\documentclass[sn-nature]{sn-jnl}

\usepackage[super]{nth}
\usepackage{graphicx}%
\usepackage{multirow}%
\usepackage{amsmath,amssymb,amsfonts}%
\usepackage{amsthm}%
\usepackage{mathrsfs}%
\usepackage[title]{appendix}%
\usepackage[table,xcdraw]{xcolor}%
\usepackage{textcomp}%
\usepackage{manyfoot}%
\usepackage{booktabs}%
\usepackage{algorithm}%
\usepackage{algorithmicx}%
\usepackage{algpseudocode}%
\usepackage{listings}%

\usepackage{tabularx}
\usepackage{longtable}
\usepackage{multirow}
\usepackage{multicol}
\usepackage{caption}
\usepackage{subcaption}
\usepackage{graphicx}
\usepackage{xspace}

\usepackage{tikz}

\usepackage{xr-hyper}
\usepackage{hyperref}
\usepackage{cleveref}

\newcommand{\COVID}{{COVID-19}\xspace}

\newcommand{\poli}{poli\xspace}

\unnumbered

\begin{document}


\title{Regional and Temporal Patterns of Partisan Polarization during the COVID-19 Pandemic in the United States and Canada}

\author*[1,3]{\fnm{Zachary} \sur{Yang}}\email{zachary.yang@mail.mcgill.ca}

\author[2]{\fnm{Anne} \sur{Imouza}} 

\author[3]{\fnm{Maximilian} \sur{Puelma Touzel}} 

\author[4]{\fnm{Cécile} \sur{Amadoro}} 

\author[4]{\fnm{Gabrielle} \sur{Desrosiers-Brisebois}} 

\author[1,3]{\fnm{Kellin} \sur{Pelrine}} 

\author[1,3]{\fnm{Sacha} \sur{Levy}} 

\author[3,4]{\fnm{Jean-Fran\c{c}ois} \sur{Godbout}} 

\author[1,3]{\fnm{Reihaneh} \sur{Rabbany}} 

\affil[1]{\orgdiv{School of Computer Science}, \orgname{McGill University}}

\affil[2]{\orgdiv{Political Science}, \orgname{McGill University}}

\affil[3]{\orgname{Montreal Institute for Learning Algorithms}}

\affil[4]{\orgdiv{Department of Political Science}, \orgname{University of Montreal}}



\abstract{

Public health measures were among the most polarizing topics debated online during the COVID-19 pandemic. Much of the discussion surrounded specific events, such as when and which particular interventions came into practise. In this work, we develop and apply an approach to measure subnational and event-driven variation of partisan polarization and explore how these dynamics varied both across and within countries. We apply our measure to a dataset of over 50 million tweets posted during late 2020, a salient period of polarizing discourse in the early phase of the pandemic. In particular, we examine regional variations in both the United States and Canada, focusing on three specific health interventions: lockdowns, masks, and vaccines. We find that more politically conservative regions had higher levels of partisan polarization in both countries, especially in the US where a strong negative correlation exists between regional vaccination rates and degree of polarization in vaccine related discussions. We then analyze the timing, context, and profile of spikes in polarization, linking them to specific events discussed on social media across different regions in both countries. These typically last only a few days in duration, suggesting that online discussions reflect and could even drive changes in public opinion, which in the context of pandemic response impacts public health outcomes across different regions and over time.

}

 
\maketitle

\section{Introduction}\label{sec:main}


Partisan polarization is increasingly prevalent in democracies around the world \citep{Hart2020}. In the United States, the level of opposition between Democrats and Republicans has been steadily growing for decades \citep{Iyengar2019,Pew2014} and reached unprecedented heights during the 2020 presidential election \citep{Hart2020,amlani2021partisanship}. This polarization even affected how individuals responded to the \COVID pandemic by influencing their assessment of the dangers posed by  the virus and their response to public health measures \citep{stewart2020polarization, chu2021not,gadarian2021partisanship}. Several studies have now confirmed that supporters of the Democratic party were more likely to follow social distancing measures \citep{partisan_difference_in_social_distancing,gollwitzer2020partisan,allcott2020polarization}, wear masks \citep{Kahane2021PoliticizingTM,milosh2021unmasking}, and get vaccinated \cite{covidVaccineHesitancy,druckman2021affective, ojea2022polarization} when compared to their Republican counterparts. This polarizing trend among partisans is not limited to the United States \citep{WARD2020113414,jiang2021social}. Other countries, like Canada, have also experienced the rapid politicization of pandemic responses, where researchers found that supporters of the Liberal Party were more likely to follow \COVID guidelines than supporters of the Conservative Party or the populist People's Party \citep{medeiros_gravelle_2023,pammett2022}.


While countries like the United States and Canada adopted strategies to coordinate efforts in addressing the pandemic at the national level, there was still significant variation in both the amount and types of public health interventions introduced by subnational governments. For instance, Canadian provinces implemented different policies, ranging from comprehensive lockdowns and school closures to more targeted guidelines focusing on specific populations \citep{blouin2021translating}. Similar variation can be observed in the United States, where some states implemented strict lockdown orders and mask mandates, while others refused to limit social distancing \citep{adolph2021pandemic}. Much like at the national level, these different regional policies also became rapidly politicized along partisan lines \citep{jiang2021social}, especially on social media platforms, where the politicization of the \COVID pandemic largely unfolded \citep{ashokkumar2021social,wu2021partisan, allcott2020polarization,Bridgman2020}. 


Numerous studies have now confirmed that online discussions surrounding the pandemic \citep{jiang2020political,gollwitzer2020partisan,gallotti2020assessing,lang2021maskon} exhibited clear regional patterns characterized by the same partisan animosity that impacted the heterogeneous implementation of public health measures \citep{jiang2020political} and the resulting epidemiological outcomes \citep{gollwitzer2020partisan}. Additional research highlights that these partisan divisions also contributed to the increasing polarization observed on social media \citep{jiang2020political,lang2021maskon,Bridgman2020}. The intense reactions from both supporters and opponents of public health measures \citep{rodriguez2022morbid} implies that public opinion could have significantly been influenced by local political dynamics and the geography of the pandemic \citep{clinton2021partisan}. This regional heterogeneity provides us with a unique opportunity to study polarization around specific events, topics, and regions, to understand how various factors affected compliance to COVID-19 guidelines. However, reliably measuring the polarization of public discourse at more fine-grained resolution is a challenge, even with the large quantity of human-generated text with extensive meta-data available on digital platforms.

In this work, we propose a solution to this measurement problem by introducing a comprehensive approach to better understand the geographic and event-driven variation of online partisan polarization of \COVID discussions within American states and Canadian provinces. We examine variations in public discourse as contained on Twitter (now X) to determine how polarization is related to: (1) the ideological leanings of different regions; (2) the amount of conspiracy theory-related messages that users have been exposed to; and (3) vaccination data.


The paper is organized as follows. First, we briefly outline our approach to region- and time-resolved polarization data from Twitter (X). In this section, we also describe our machine learning method to classify users as conservative or liberal and justify our choice of topic-conditioned language dissimilarity as a proxy for partisan polarization. Next, we present the results of applying our approach to a large-scale dataset we collected in 2020, filtered through three prominent pandemic-related public health interventions: lockdowns, masks, and vaccines. Our findings indicate that conservative regions in both countries exhibited higher polarization levels on these topics overall. We also find strong negative correlation between vaccination rates in different U.S. regions and the level of polarization in their online discussions related to vaccines. We close with a discussion of limitations of the approach and promising new areas of application.




\section{Approach} 

\begin{figure}[ht!]
 \centering
 \resizebox{\linewidth}{!}{

\includegraphics[width=\textwidth]{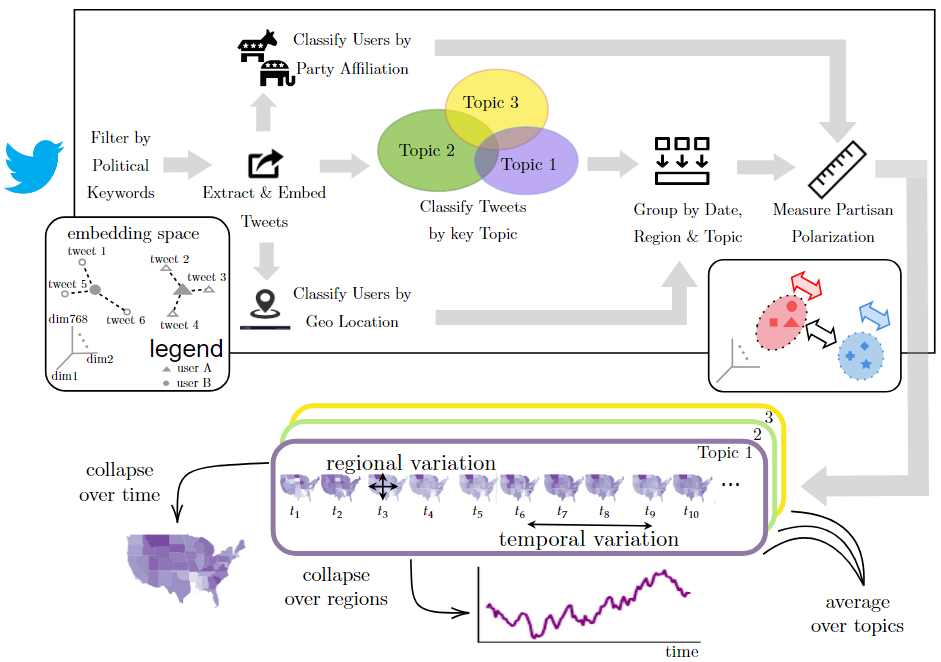}
}
 \caption{Overview of the proposed method to estimate partisan polarization over date, region, and topic (top), as well as how to analyze this data by collapsing it over any of those three dimensions (bottom). We studied the topics of lockdowns, masks, and vaccines.}
 \label{fig:Methodology_Flow}
\end{figure}
We developed a method to measure geographically-resolved partisan polarization over time from large-scale social media message datasets (see \cref{fig:Methodology_Flow}). 
%
The language of political discussions across socio-demographic groups can vary significantly \citep{Diermeier2012}, each having their own lexicon, so dissimilar language on its own does not imply polarized positions. However, when conditioning on discussion of the same, contentious topic, the dissimilarity of the language used by different demographics is more likely to reflect alternative semantic understandings of that specific topic, which we assume \textit{are} highly correlated with polarization. This correlation is weakened by linguistic differences not captured by the particular definition of semantic separation used, so the latter is only a noisy proxy of polarization strength. That said, we expect that a stronger correlation between semantic separation and polarization exists when language is represented in a more expressive model. Inspired by the demonstrated capacity of modern vector embeddings to represent the semantics of words, our approach focuses on transforming the sentences of social media posts using RoBERTa, a powerful open-source, language-embedding model \citep{liu2019roberta}. As an indicator for polarization, we then measure dissimilarity by how far apart the tweets of left and right-leaning users are in this embedding space. In particular, we use the C-index \cite{Hubert1976AGS}, a robust clustering measure based on the average of pairwise distances of embedded partisan users within a partisan group relative to the average of the largest and smallest distances overall. To label partisans in our data, we developed and validated a machine learning method that identifies users as conservative or liberal on the ideological spectrum by cross-indexing multiple metadata sources. We also developed and validated a method to geolocate users to resolve polarization's geographic heterogeneity. Details of these components of the approach can be found in the Methods section.

\subsection{Application to late 2020 pandemic discourse in the United States \& Canada}
We collected a large-scale dataset of \COVID political discussions on Twitter (X) occurring between October \nth{9}, 2020, to January \nth{4}, 2021, comprising about 46.6 million tweets linked to Canada and 12.5 million tweets linked to the United States.
We geolocated users based on their provided location and classified them by their declared party affiliation. Specifically, we include identifiers for the two major liberal and conservative ideological divisions in each country: the left (Liberal Party, New Democratic Party, and Green Party) and the right (Conservative Party and People's Party) for Canada; and the Democratic (left) and Republican (right) parties for the United States.
Using official population census and election results, we verify that these data provide a politically balanced set of users in the different regions of these two countries.
For each of the users, we then compute a vector representation for the language they used in their social media messages during this period. We condition on region, time, and topic and for each combination compute the value of the C-index as a proxy for polarization. 
%
%
%
To narrow the content of the messages analyzed, we focus on three specific topics of discussion: lockdowns, masks and vaccines. These topics were chosen because of their salience for polarized discourse around the pandemic  \citep{lang2021maskon,Cascini2022,Wicke2020}, and because they span different types of interventions (group behaviour, individual behaviour, and medicine, respectively).
Based on these measurements, we then compare the polarization observed in different American states and Canadian provinces over time for each of the three topics. We also look into how polarization is correlated with epidemiological data and conspiracy-related content. We refer the reader to the Methods section for further details. 

\section{Results}
We organize the presentation of results as follows. First, we report the geographical trends of the observed partisan polarization in the United States and Canada and confirm that conservative states and provinces display more polarized online discourse. Next, we highlight the correlation between partisan polarization on the topic of vaccines and the vaccination rates found across U.S. states. We then present our event-based analysis of the temporal patterns of polarization at the national level in both countries and report correlations between polarization and vaccination data, as well as the volume of conspiracy-related content on Twitter (X). Finally, we examine the different peaks in polarization and explain how they relate to various polarizing events.  


\begin{figure}[tbhp!]\centering
    \begin{subfigure}{0.47\linewidth}
        \centering
        \includegraphics[scale=0.6,trim={0 0 0 1cm},clip]{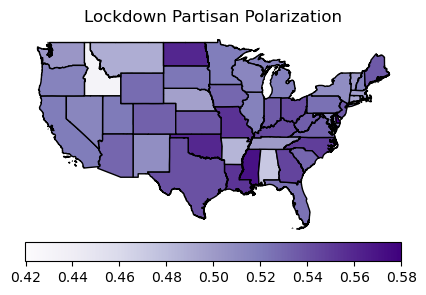}
        \caption{Lockdown Polarization}
        \label{fig:US_AVG_WKLY_LCKDWN_POL}
    \end{subfigure}
    \hspace{0.01\linewidth}
    \begin{subfigure}{0.47\linewidth}
        \centering
        \includegraphics[scale=0.6,trim={0 0 0 1cm},clip]{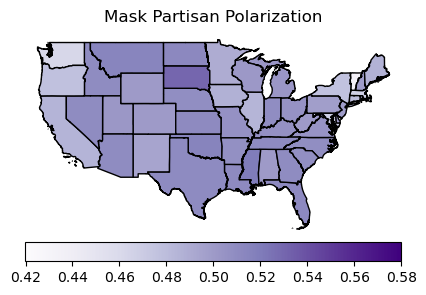}
        \caption{Mask Polarization}
        \label{fig:US_AVG_WKLY_MASK_POL}
    \end{subfigure}
    \\
    \begin{subfigure}{0.47\linewidth}
        \centering
        \includegraphics[scale=0.6,trim={0 0 0 1cm},clip]{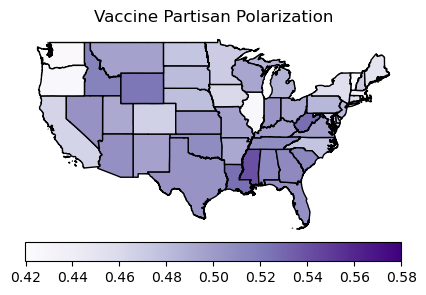}
        \caption{Vaccine Polarization}
        \label{fig:US_AVG_WKLY_VAX_POL}
    \end{subfigure}    
    \hspace{0.01\linewidth}
     \begin{subfigure}{0.47\linewidth}
        \centering
        \includegraphics[scale=0.6,trim={0 0 0 1cm},clip]{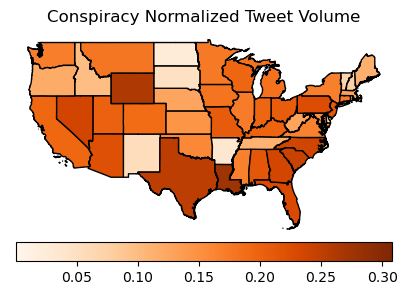}
        \caption{\% of Conspiracy-related Tweets}
    \label{fig:US_AVG_WKLY_CONSP_POL}
    \end{subfigure}
    \caption{Regional distribution of partisan polarization in the United States on three key topics of Lockdown (a), Mask (b), and Vaccines (c). Color intensity from light to dark gives the amount of polarization measured weekly between October \nth{11}, 2020 to January \nth{3}, 2021 and then averaged over the 12 weeks. We also report the average weekly percentage of conspiracy-related tweets that are posted from users in each region in panel (d).}
    \label{fig:US_AVG_WKLY_POL}
\end{figure}

\begin{figure}[h!]\centering
    \begin{subfigure}{0.47\linewidth}
        \centering
        \includegraphics[scale=0.5,trim={0 0 0 1cm},clip]{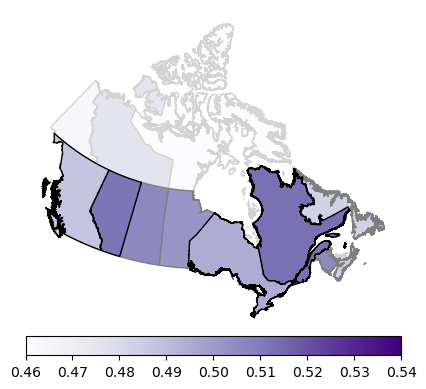}
        \caption{Lockdowns Polarization}
        \label{fig:CAN_AVG_WKLY_LCKDWN_POL}
    \end{subfigure}
    \hspace{0.01\linewidth}
    \begin{subfigure}{0.47\linewidth}
        \centering
        \includegraphics[scale=0.5,trim={0 0 0 1cm},clip]{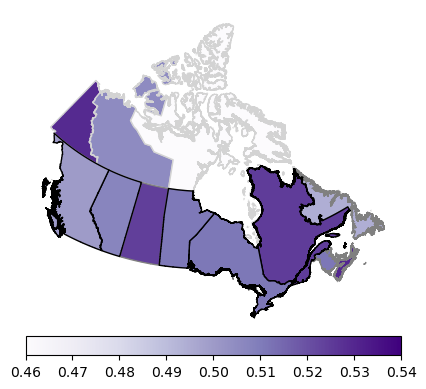}
        \caption{Masks Polarization}
        \label{fig:CAN_AVG_WKLY_MASK_POL}
    \end{subfigure}
    \\
    \begin{subfigure}{0.47\linewidth}
        \centering
        \includegraphics[scale=0.5,trim={0 0 0 1cm},clip]{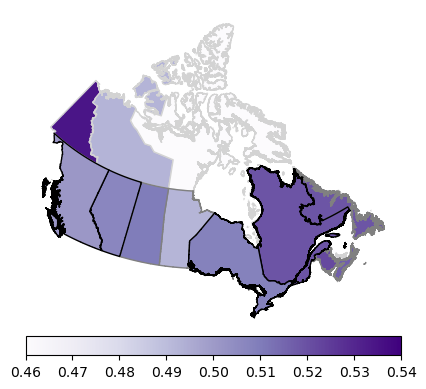}
        \caption{Vaccines Polarization}
        \label{fig:CAN_AVG_WKLY_VAX_POL}
    \end{subfigure}    
    \hspace{0.01\linewidth}
    \begin{subfigure}{0.47\linewidth}
        \centering
        \includegraphics[scale=0.5,trim={0 0 0 1cm},clip]{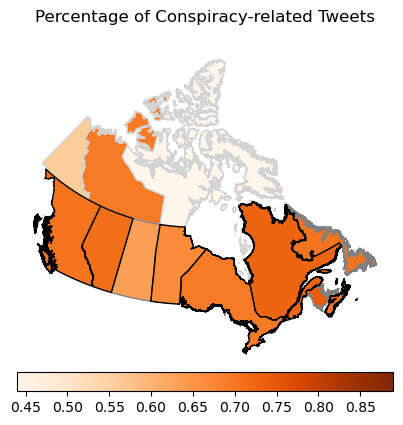}
        \caption{\% of Conspiracy-related Tweets}
        \label{fig:CAN_AVG_WKLY_CONSP_POL}
    \end{subfigure}
    \caption{
    Regional distribution of partisan polarization in Canada on three key topics of Lockdowns (a), Masks (b), and Vaccines (c). 
    The polarization is measured weekly between October \nth{11}, 2020 to January \nth{3}, 2021 and the averaged over 12 weeks is used for this plot. We also report the average weekly percentage of conspiracy-related tweets that are posted from users in each region (d). Provinces and territory boundaries are colored based on the number of users we had in our data from those regions, which indicates the support for our measurement: Light-grey for less than 100 users, grey for between 100 and 1,000 users and black for greater than 1,000 users.}
    \label{fig:CAN_AVG_WKLY_POL}
\end{figure}

\subsection{Regional Variation in Partisan Polarization} 

Our analysis begins by visualizing the geography of partisan polarization in \cref{fig:US_AVG_WKLY_POL} and \cref{fig:CAN_AVG_WKLY_POL} for the United States and Canada. The regional heterogeneity in the amount of polarization observed over different topics is apparent in both countries. We also see heterogeneity in the amount of conspiracy-related tweets shown in \cref{fig:US_AVG_WKLY_POL}d and \cref{fig:CAN_AVG_WKLY_POL}d for both countries, respectively.

\begin{figure}[tbph!]
    \centering
    \includegraphics[scale=0.75]{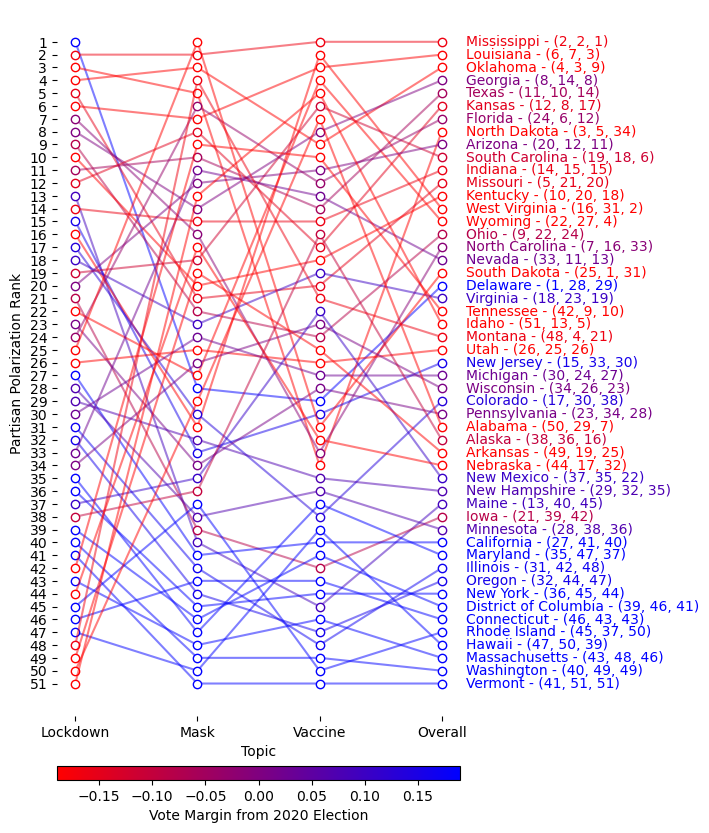}
    \caption{Ranking of American states partisan polarization per topic and overall. Ranking of 1 signifies the highest average weekly polarization between October \nth{11}, 2020 to January \nth{3}, 2021 (12 weeks). State names are colored based on the vote margin for the conservative party from the 2020 United States Presidential Election (Conservative Party: Republican Party; Liberal Party: Democratic Party). 
    }
    \label{fig:us_state_total_weekly_pol_election_upa}
\end{figure}
\begin{figure}[h!]
    \centering
    \includegraphics[scale=0.7]{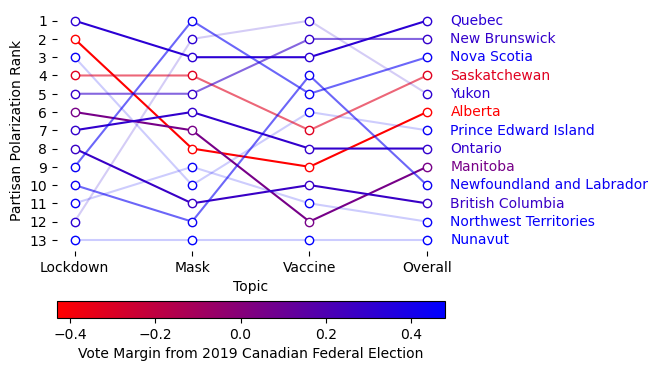}
    \caption{Partisan polarization ranking of Canadian provinces and territories per topic and overall. A ranking of 1 signifies the highest average weekly polarization between October \nth{11}, 2020 to January \nth{3}, 2021 (12 weeks). Province or territory names are colored (red to blue) based on the vote margin for the conservative party family from Canada’s 2019 Federal Election (Liberal Party Family: Liberal, New Democratic Party, Green; Conservative Party Family: Conservative, People’s Party). Line colors have a transparency to reflect the support for the measurement, based on the number of users in that region.  
    }
    \label{fig:cad_state_total_weekly_pol_election_upa}
\end{figure}

\begin{figure}[tbph!]
    \centering

    \begin{subfigure}{\linewidth}
        \centering
        \includegraphics[width=\linewidth]{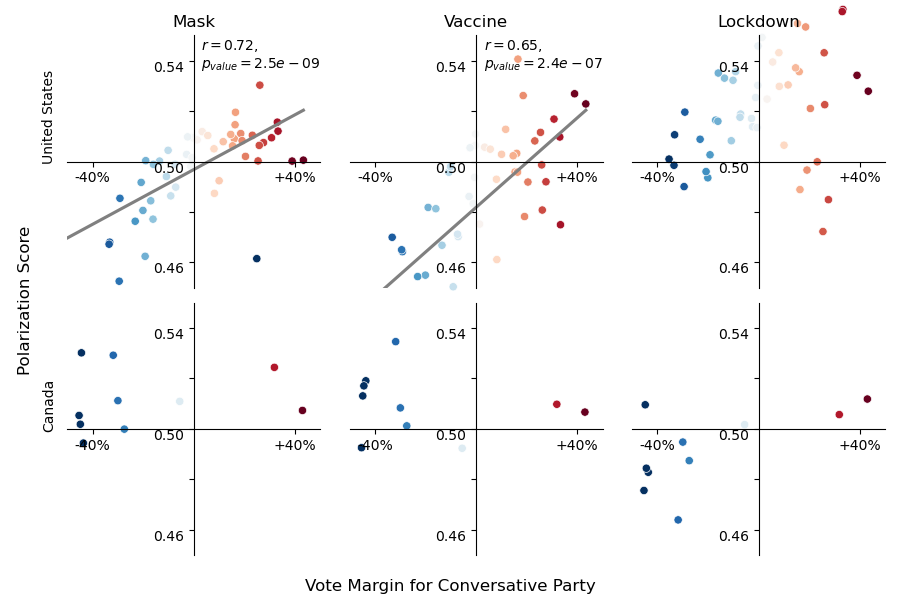}
        \caption{Correlation between Polarization Score and Vote Margin for Conservative Party. Colors (blue to red) are conservative party vote margin (same as \cref{fig:us_state_total_weekly_pol_election_upa} and \cref{fig:cad_state_total_weekly_pol_election_upa}). Significant correlation between Polarization Score and Vote Margin is found for the US discourse on masks and on vaccines for which the respective Pearson $r$ correlation and $p$-value is shown.}
        \label{fig:corr_US}
    \end{subfigure}
    \\
    \begin{subfigure}{\linewidth}
        \centering
         \includegraphics[scale=0.55]{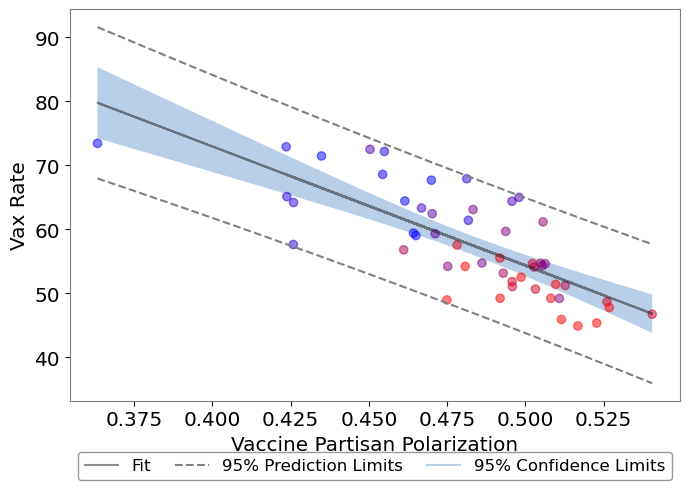}
        \caption{Relation between vaccines polarization and vaccination rates in the  United States. Color (blue to red) is again the respective conservative party vote margin from the 2020 U.S. Presidential Election. The correlation is \textbf{-0.77} with CI = [-0.86, -0.62] (n = 51,  p = 6.97e-11).}
        \label{fig:corr_vax}
    \end{subfigure}
    \caption{}
\end{figure}

Next, we analyze how this heterogeneity varies with the partisan leanings found in each region by analyzing election voting patterns (the 2020 presidential election in the case of the United States and the 2019 federal election for Canada). 
Our first observation is that \textit{conservative states and provinces show higher levels of polarization compared to their liberal counterparts}. To display results over all covered regions, we show the polarization rankings for American states and Canadian provinces in  \cref{fig:us_state_total_weekly_pol_election_upa} and \cref{fig:cad_state_total_weekly_pol_election_upa}, respectively. This ranking is applied separately to each of the three topics and is based on their weekly polarization averaged over the 12-week period. An additional fourth ranking labelled \textit{overall} is shown and gives the average over the three topics. Each region is associated with a color graded from blue to red based on the vote margin for the Republican party (US) or the conservative party family (Canada) obtained from the votes reported in the most recent election in their corresponding country. In these figures,  a blue to red color gradient for conservative to progressive is used such that the names of predominantly conservative/Republican regions appear in red, predominantly liberal/Democratic regions in blue, and mixed or less definitive regions in purple. 
%
%
Referring to \cref{fig:us_state_total_weekly_pol_election_upa}, in the United States we can see clearly that conservative states are more polarized compared to liberal states overall and specifically on discussions related to masks and vaccines. The ranking is more mixed in the discussions about lockdown measures with outliers from both liberal and conservative states; namely Idaho, Alabama, and Arkansas showing the least polarization, and Delaware (ranked 1st), Colorado (ranked 17th), and New Jersey (ranked 15th) showing higher values. The expected relationship between pandemic response and state partisanship is however still present for lockdown discussions, with liberal states such as Vermont (ranked 41st) and Massachusetts (ranked 43rd) displaying less polarization compared to more conservative states such as Mississippi (ranked 2nd), North Dakota (ranked 3rd), and Oklahoma (ranked 4th).

Looking now at Canada \cref{fig:us_state_total_weekly_pol_election_upa}, we find that Alberta, a conservative province, shows higher polarization compared to Ontario, and British Columbia (among the Canadian provinces with the most number of social media users). Quebec is overall the highest ranked province. Although the pandemic was highly polarized in Quebec---e.g., with violent protests \cite{Rowe2020}---we want to acknowledge the limitation of our study, which was focused on the English language; in Quebec, the main language is French whereas only English tweets were included in our analysis. 

Finally, the correlation between polarization and the partisan vote margin is more clearly represented in the scatter between the two, shown for both countries in \cref{fig:corr_US}. We see a strong and significant correlation between Republican vote share in the United States and the polarization index around masks and vaccines discourse, but not lockdowns. The remaining associations (Lockdown for US, and all topics for Canada) are, however, insignificant (for Canada this is in part due to the relatively small number of regions). 

The strong correlation that we observe between the polarization score for discourse around vaccines in conservative-leaning states follows the well-known negative correlation between Republican vote share and vaccination rates \citep{Albrecht2022}.
Since vaccines were not available yet over this time period, we nevertheless present a comparison using official vaccination rates measured for different states by the U.S. Centers for Disease Control and Prevention for a similar period of time one year later, after the vaccines were rolled out (i.e. October \nth{11}, 2021 to January \nth{3}, 2022). Averaging on a weekly basis, we confirm this correlation in \cref{fig:corr_vax}, where we observe that \textit{vaccine polarization is strongly negatively correlated to vaccination rates in the different American states}. We did not observe a similar pattern in Canada, due to small sample size and the implementation of vaccine mandates. 
While disentangling the causal relationships among conservative vote margin, polarization score, and vaccination rate is not possible here, the results suggest that polarized discourse played a role in shaping the highly heterogeneous vaccination rates across the U.S.

\subsection{Temporal Variation in Partisan Polarization}
\begin{figure}[h!]

    \centering

    \begin{subfigure}{0.48\linewidth}
        \centering
        \includegraphics[scale=0.43,trim={0 0 1cm 0}, clip]{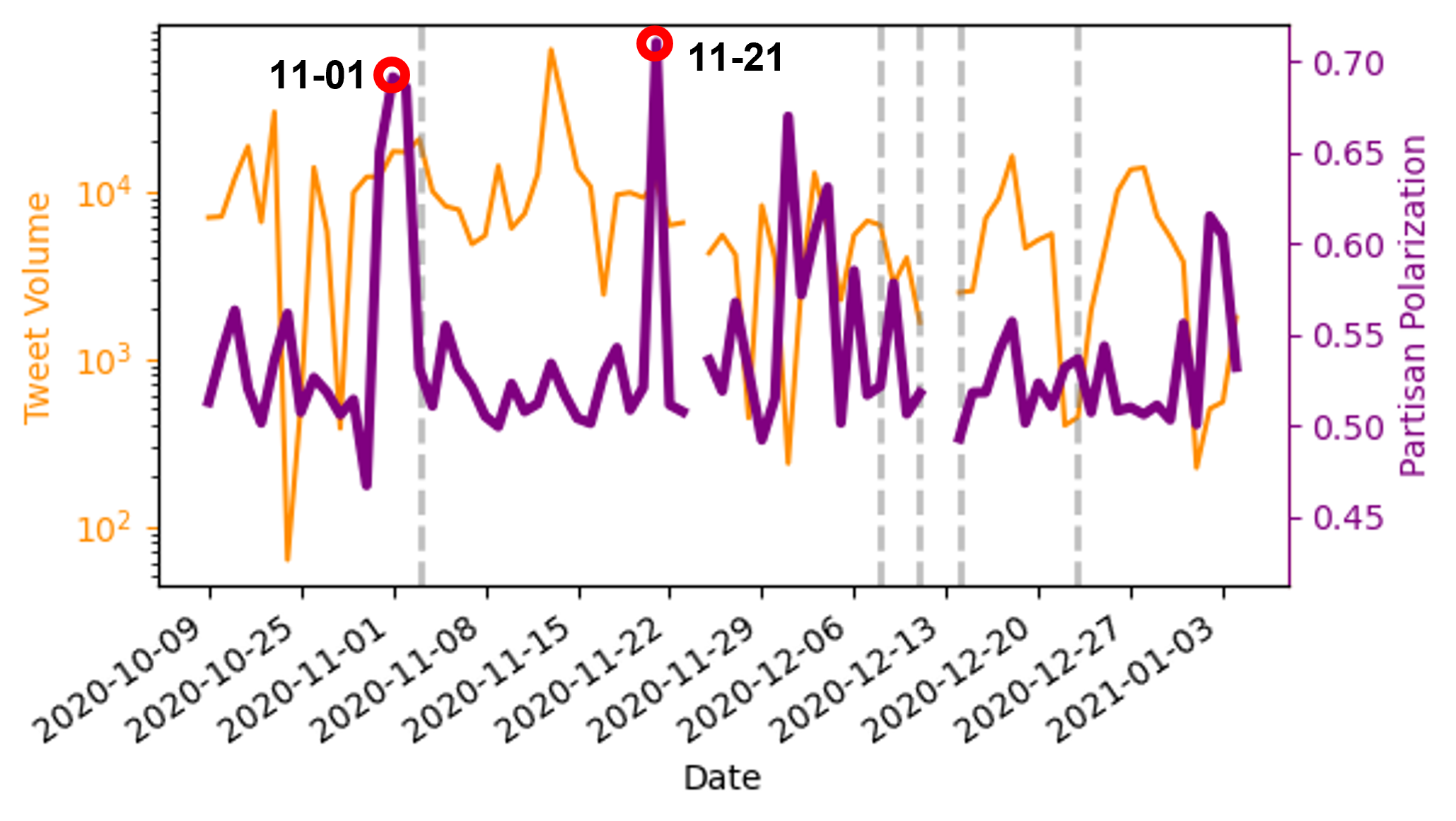}
        \caption{US Daily \textbf{Lockdown} Polarization}
        \label{fig:us_daily_lockdown_polarization}
    \end{subfigure}
    \hspace{0.02\linewidth}
    \begin{subfigure}{0.48\linewidth}
        \centering
        \includegraphics[scale=0.43,trim={0.6cm 0 0 0}, clip]{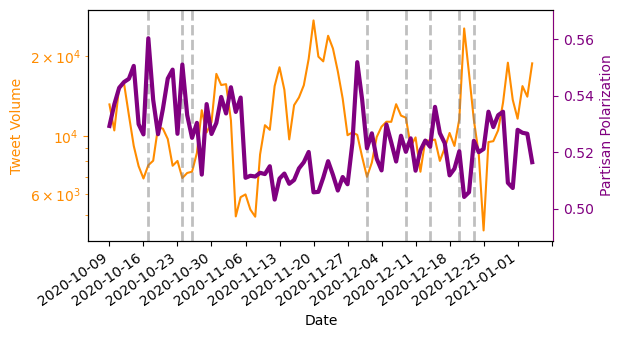}
        \caption{Canada Daily \textbf{Lockdown} Polarization}
        \label{fig:cad_political_daily_lockdown_polarization}
    \end{subfigure}
    \\
    
    \begin{subfigure}{0.48\linewidth}
        \centering
        \includegraphics[scale=0.43,trim={0 0 1cm 0}, clip]{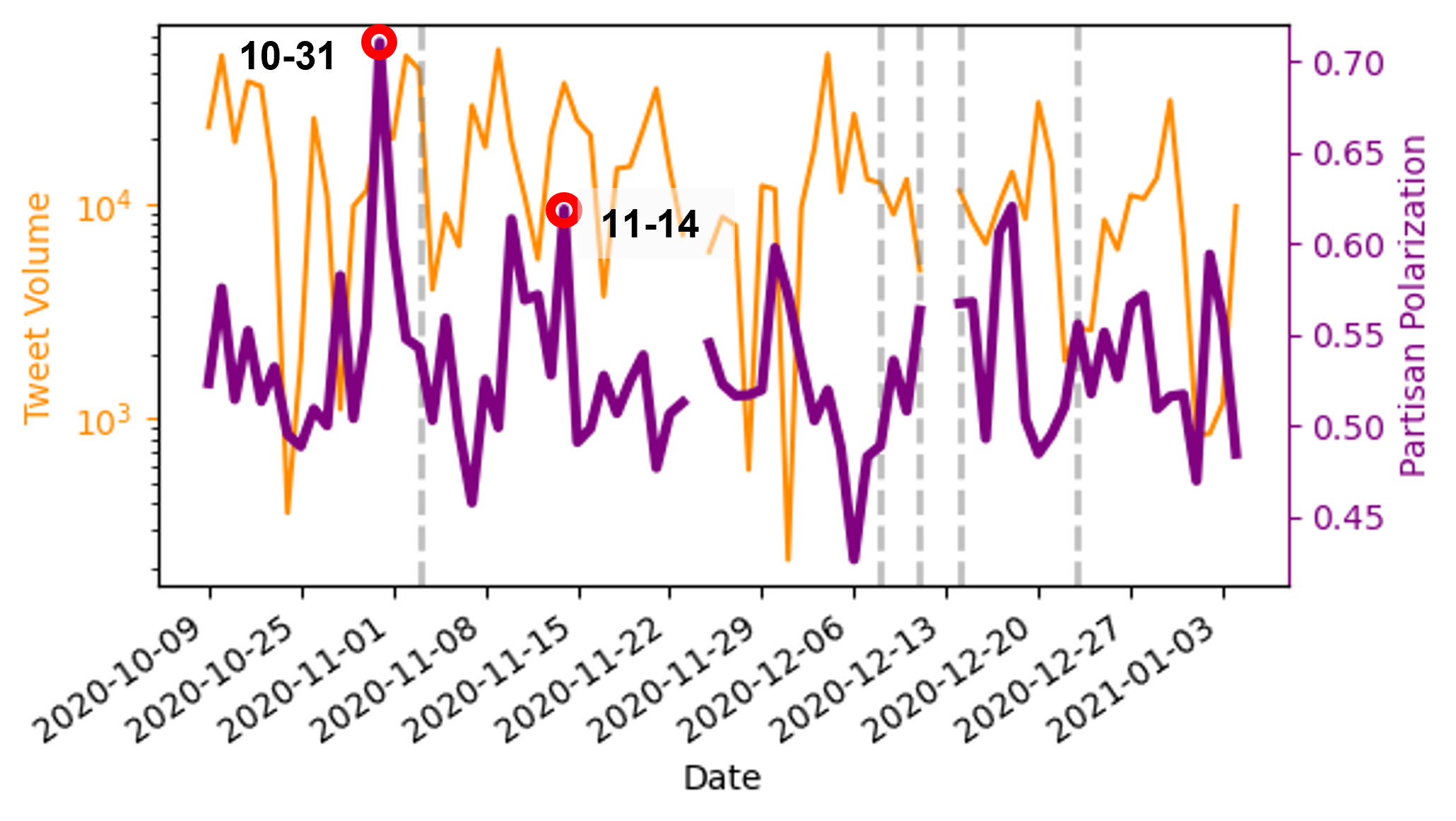}
        \caption{US Daily \textbf{Mask} Polarization}
        \label{fig:us_daily_mask_polarization}
    \end{subfigure}
    \hspace{0.02\linewidth}
    \begin{subfigure}{0.48\linewidth}
        \centering
        \includegraphics[scale=0.43,trim={0.6cm 0 0 0}, clip]{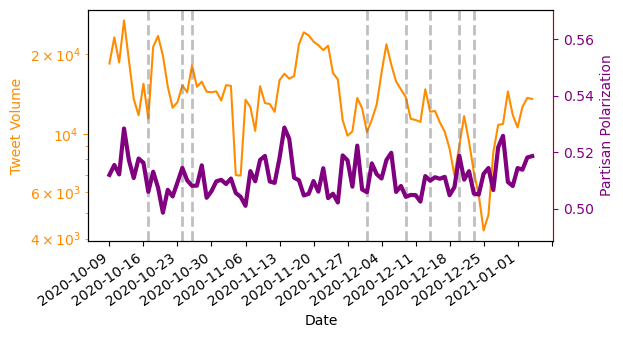}
        \caption{Canada Daily \textbf{Masks} Polarization}
        \label{fig:cad_political_daily_mask_polarization}
    \end{subfigure}
    \\
    
    \begin{subfigure}{0.48\linewidth}
        \centering
        \includegraphics[scale=0.43,trim={0 0 1cm 0}, clip]{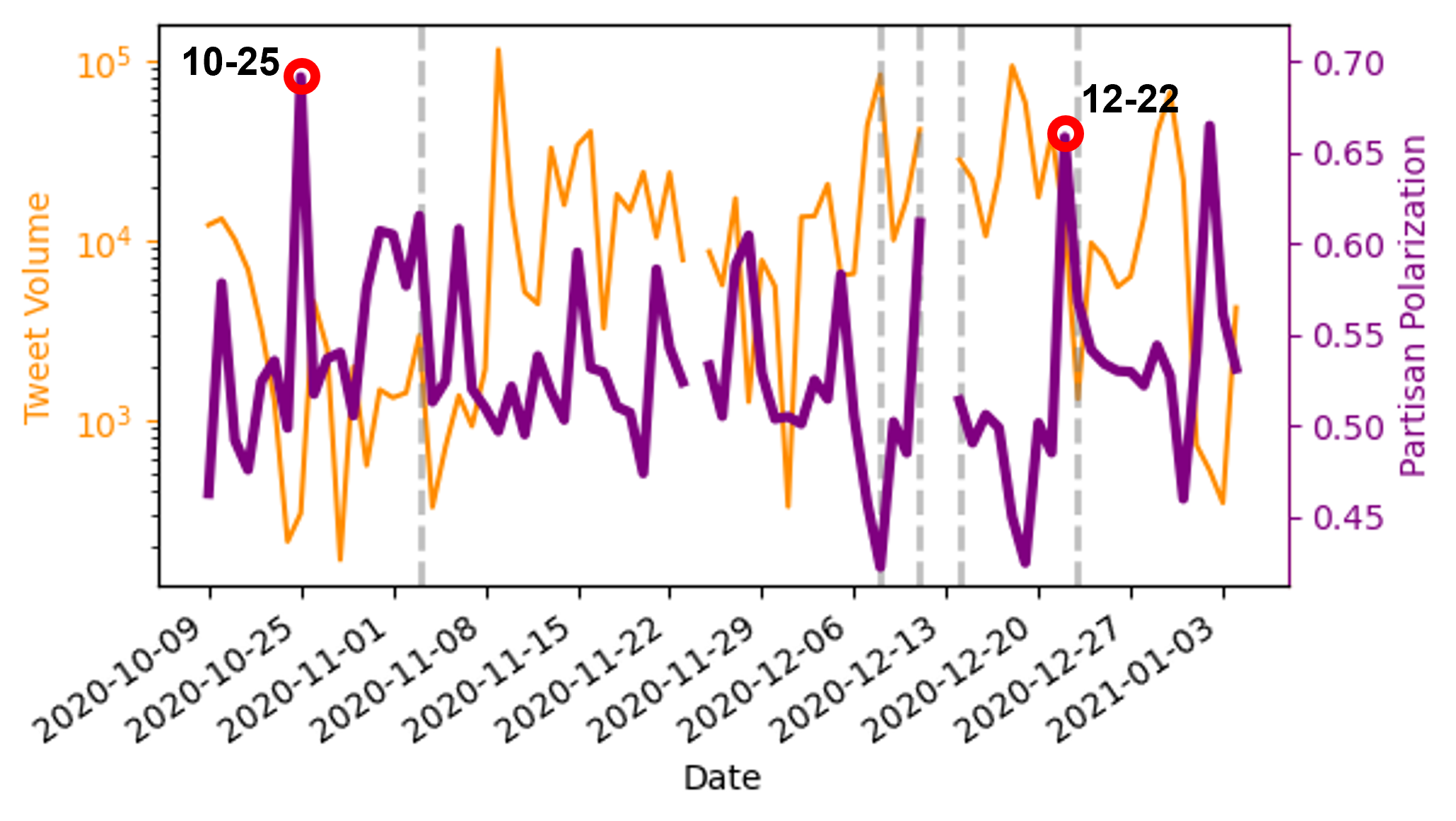}
        \caption{US Daily \textbf{Vaccine} Polarization}
        \label{fig:us_daily_vaccine_polarization}
    \end{subfigure}
    \hspace{0.01\linewidth}
    \begin{subfigure}{0.48\linewidth}
        \centering
        \includegraphics[scale=0.43,trim={0.5cm 0 0 0}, clip]{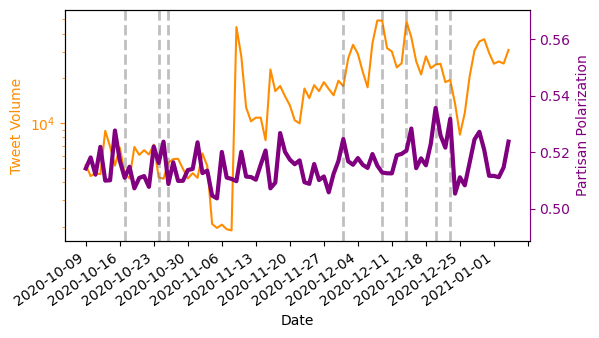}
        \caption{Canada Daily \textbf{Vaccine} Polarization}
        \label{fig:cad_political_daily_vaccine_polarization}
    \end{subfigure}
    
    \caption{
    Daily trends of partisan polarization in the\textbf{ United States } and \textbf{ Canada } from October \nth{9}, 2020 to January \nth{3}, 2021. The vertical dashed lines denote pre-selected political and vaccine-related events as explained in the text. In addition to the polarization measure (purple line), we also report the tweet volume, in log-scale, on the corresponding topic (yellow line) per day which denotes the size of support for our measurement.
    }

    \label{fig:us_cad_political_daily_pol}
\end{figure}

We next focus on the temporal trends of daily partisan polarization at the national level for each topic and overall, as displayed in \cref{fig:us_cad_political_daily_pol} for the United States and Canada, respectively. In these figures, the value of the metric fluctuates rapidly on the timescale of days. This is on the faster end of the range of timescales found in other topic tracking studies, e.g., \cite{leskovec2009}. These short timescales are consistent with our assumption that language adapts quickly in rapid anticipation of or as an immediate response to specific events. In particular, we considered two kinds of events. First, we preselected political and vaccine-related events (shown in \cref{table:events} and as vertical lines in  \cref{fig:us_cad_political_daily_pol}). These provide the scaffold for the socio-political trajectory of each country related to political discourse and pandemic response. Second, we detected highly polarized events through analysis of the highest two peaks in polarization (shown in \cref{table:peak_events} and as red circles in \cref{fig:us_cad_political_daily_pol}). We also show in the figures the tweet volume as a relative indicator of day-by-day reliability of the estimation of the polarization score.
While we do not have direct causal evidence linking a highlighted peak in the polarization score to a specific event, we do find that many of \textit{the largest polarization peaks occur around highly contentious events} related to each country's specific context (\cref{table:peak_events}). In the following two sections, we summarize these events and discuss how they relate to the topic for which the polarization simultaneously peaks.

\begin{table}[t!]
\centering

\begin{subtable}{\linewidth}
\centering
\captionsetup{width=\linewidth}
\caption{Major political and pandemic-related events in each country such as wheen the FDA (U.S. Food \& Drug Administration), and PHAC (Public Health Agency of Canada) approved vaccines. } 
\label{table:events}
\begin{tabular}{l|ll}
Date                                           & Event                                                 & Country                        \\ \hline
Nov. 3                                         & US National Election                                  & US                             \\
\rowcolor[HTML]{EFEFEF} 
Oct. 24                                        & BC General Election                                   & Canada                         \\
\rowcolor[HTML]{EFEFEF} 
Oct. 26                                        & Saskatchewan General Election                         & Canada                         \\
Dec. 8                                         & States resolve controversies                          & US                             \\
\rowcolor[HTML]{EFEFEF} 
Dec. 9                                         & PHAC approves Pfizer vaccine                          & Canada                         \\
Dec. 11                                        & FDA approves vaccines                                 & US                             \\
\multicolumn{1}{c|}{}                          & Electoral votes submitted                             & US                             \\
\multicolumn{1}{c|}{\multirow{-2}{*}{Dec. 14}} & \cellcolor[HTML]{EFEFEF}Vaccination begins            & \cellcolor[HTML]{EFEFEF}Canada \\
Dec. 20                                        & Moderna vaccine distributed                           & US                             \\
                                               & Election votes arrive                                 & US                             \\
\multirow{-2}{*}{Dec. 23}                      & \cellcolor[HTML]{EFEFEF}PHAC approves Moderna Vaccine & \cellcolor[HTML]{EFEFEF}Canada
\end{tabular}
\end{subtable}
\\

\begin{subtable}{\linewidth}
\centering
\captionsetup{width=\linewidth}
\caption{Polarization peaks and their corresponding events. For each topic, we analyzed the two highest peaks, and inferred the content discussed on those peaks.} 
\label{table:peak_events}
\footnotesize
\begin{tabular}{l|ll|ll}
                                                & \multicolumn{2}{c|}{United States}    & \multicolumn{2}{c}{Canada}              \\
\multirow{-2}{*}{Topic}                         & Date    & Polarizing Event            & Date    & Polarizing Event              \\ \hline
                                                & Nov. 1  & Viral tweet by Trump        & Oct. 17 & Toronto Mask Measures Protest \\
\multirow{-2}{*}{Lockdown}                      & Nov. 21 & Unidentified topic          & Oct. 29 & Calgary Mask Measures Protest \\
\rowcolor[HTML]{EFEFEF} 
\cellcolor[HTML]{EFEFEF}                        & Oct. 21 & Viral tweet by Trump        & Oct. 12 & Unidentifed topic             \\
\rowcolor[HTML]{EFEFEF} 
\multirow{-2}{*}{\cellcolor[HTML]{EFEFEF}Masks} & Nov. 14 & Biden proposes mandates     & Nov. 14 & PHAC recommends masks         \\
                                                & Dec. 20 & Moderna vaccine distributed & Dec. 20 & Moderna distributed in U.S.   \\
\multirow{-2}{*}{Vaccines}                      & Dec 22  & Biden gets vaccinated       & Dec. 23 & PHAC approves Moderna        
\end{tabular}
\end{subtable}
\caption{Major dates and peaks within the United States and Canada during 2020.}
\label{tab:table1}
\end{table}

\paragraph{United States Polarization Timecourse}
The left column of \cref{fig:us_cad_political_daily_pol} reports the daily polarization measured for the three key topics of lockdowns, masks, and vaccines. 
%
Peaks in polarization on the lockdown topic in \cref{fig:us_daily_lockdown_polarization} may correspond to partisan differences in public support (or discontent) and discourse surrounding \COVID measures. In the days leading up to the 2020 Presidential Election on November 3rd, a pillar of President Trump's campaign messaging on the pandemic characterized lockdowns as tyranny and economic repression \citep{Algara2022}. For example, on \textbf{November \nth{1}, 2020}, the date of the second-largest peak, Trump made a highly controversial claim by stating that the election was a choice between implementing deadly lockdown measures supported by Biden or an efficient end to the \COVID crisis with a safe vaccine \citep{Bryant2020}. Trump also made other similar claims on Twitter (X) during this period, e.g., when he said (sic): ``Biden wants to LOCK DOWN our Country, maybe for years. Crazy! There will be NO LOCKDOWNS. The great American Comeback is underway!!!'' \citep{Algara2022}. 

%
%

Next, contentious debates related to masks were found coincident with peaks in polarization, as shown in \cref{fig:us_daily_mask_polarization}. 
For example, while we did not find an event external to social media on \textbf{October \nth{31}, 2020}, the date of the highest peak, we did find that the most retweeted tweet by Democrats on that day was ``RT @JoeBiden: Be a patriot. Wear a mask.''. This, in turn, generated strong responses that day from Republicans with the third most retweeted tweet within this group: ``RT @RealBrysonGray: There's literally nothing patriotic about being so scared of a virus with a 99.9...''. This is then possibly an example of influencer post-driven, rather than real world event-driven polarization. 
Another set of divisive messages were observed on \textbf{November \nth{14}, 2020}, the next highest peak, after presidential candidate Biden proposed mandatory mask mandates, and South Dakota Governor Kristi Noem announced her opposition to this measure;  nearly half of all the top retweets referred to Noem's statement. 

%

Finally, \cref{fig:us_daily_vaccine_polarization} reports trends in polarization around the topic of vaccines. Here, some of the peaks observed are simultaneous with important events surrounding \COVID vaccine efficacy. 
We see that the second-largest peak occurred on \textbf{December \nth{22}, 2020}, a day after Biden received his first \COVID vaccine shot \citep{Higgings2020}. 
The most retweeted tweet for both partisan groups that day was ``RT @JoeBiden: Today, I received the COVID-19 vaccine. To the scientists and researchers who worked tirelessly to make this possible - than…". However, while supporters of Biden congratulated him, by tweeting messages like ``RT @YAFBiden: And just like that, @JoeBiden has received the COVID-19 vaccine!'', opponents instead promoted pro-Trump messages, \textit{e.g.} ``RT @TheLeoTerrell: Finally a @JoeBiden confession. He finally gave credit to @realDonaldTrump and \#OperationWarpSpeed. It's about time.''. 

\paragraph{Canadian Polarization Timecourse}
The right column of \cref{fig:us_cad_political_daily_pol} shows the daily polarization measured for the three key topics of lockdowns, masks, and vaccines. 
The pre-selected events in the Canadian timeline (\cref{table:events}) are marked as vertical lines in the figure. 
%
%

In \cref{fig:cad_political_daily_lockdown_polarization}, on the topic of lockdowns, we observe the highest peak on \textbf{October \nth{17}, 2020}, which coincides with the Toronto anti-mask protest, a large demonstration where thousands of protesters rallied against COVID-19 lockdown measures. 
The second-highest peak is observed on \textbf{November \nth{29}, 2020}, when the national news reported a Calgary Mask Measures protest on the preceding day \citep{Rieger2020}. 

The highest polarization peak on mask-related tweets is found on \textbf{November \nth{14}, 2020}, as seen in \cref{fig:cad_political_daily_mask_polarization}, coincident with the peak in the US plot \ref{fig:us_daily_mask_polarization} mentioned earlier. This event was discussed by the conservative-Party family users, showing how partisan discourse in the U.S. might be driving some polarization in Canada.


%
%
Looking at the polarization of discussions about vaccines in \cref{fig:cad_political_daily_vaccine_polarization}, we also observe the highest peaks are in response to key vaccine-related events: The two highest peaks in polarization observed on \textbf{December \nth{20} and \nth{23}, 2020} coincided with the distribution of the Moderna \COVID vaccines in the U.S. and Health Canada's approval of the Moderna vaccine \citep{HealthCanadaModerna2020}, respectively. While the former is an event associated with the United States, it led to discussions in Canada about vaccine prioritization and availability \citep{Whiteal2020}. On the \nth{20}, top retweets by liberal Party Family users focused on news of the Republican politicians being first in line for the vaccine, while conservative Party Family users retweeted more diverse anti-vaccine sentiment. On \textbf{December \nth{23}, 2020}, the top retweets were strong sentiments in support of and in opposition to the approval.
%
%

\paragraph{Aggregate Polarization}
\begin{figure}[t!]
    \centering
 \begin{subfigure}{0.48\linewidth}
        \centering
        \includegraphics[scale=0.45]{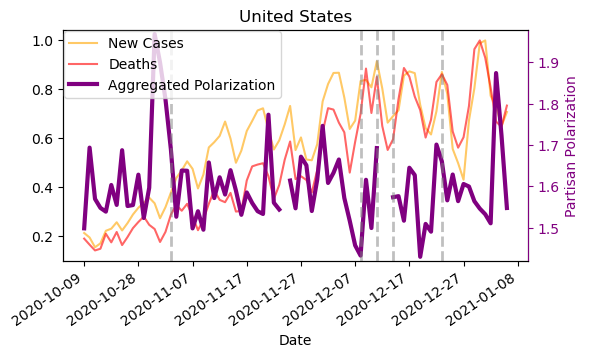}
        \caption{}
        \label{fig:us_unweighted_aggregated_poli}
    \end{subfigure}
    \hspace{0.01\linewidth}
    \begin{subfigure}{0.48\linewidth}
        \centering
        \includegraphics[scale=0.5,trim={0 0 0 0.6cm},clip]{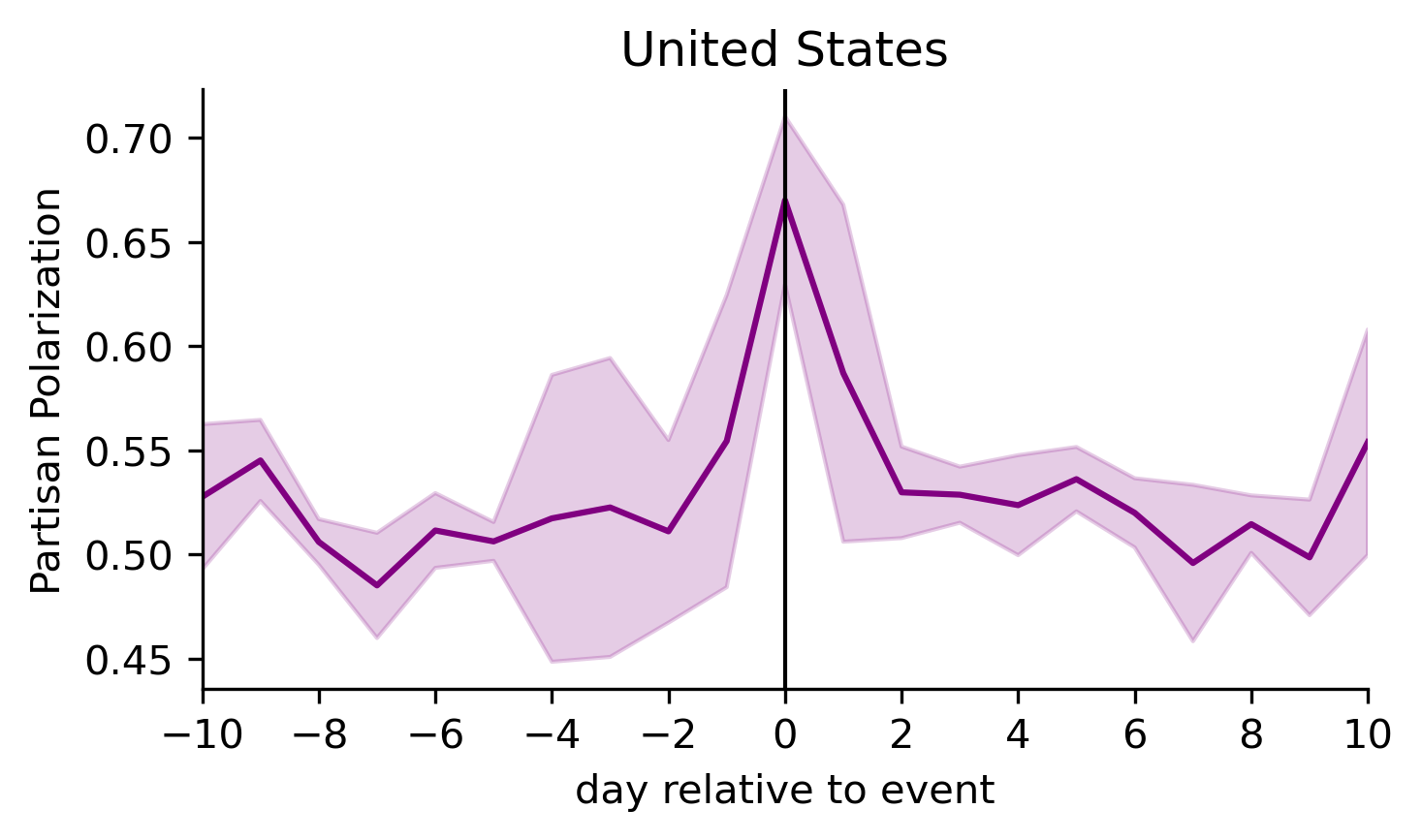}
        \caption{}
        \label{fig:Us_eventtriggered}
          \end{subfigure}
    \begin{subfigure}{0.48\linewidth}
        \centering
        \includegraphics[scale=0.45]{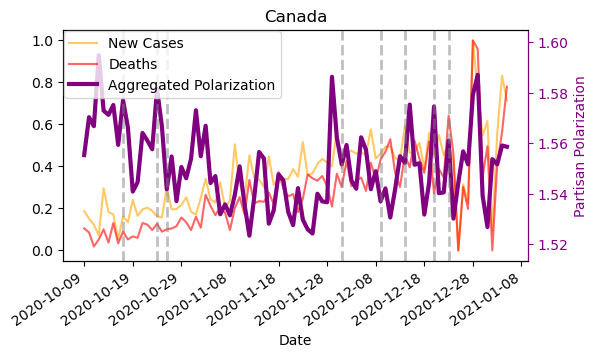}
        \caption{}
        \label{fig:cad_political_unweighted_aggregated_poli}
    \end{subfigure}
    \hspace{0.01\linewidth}
    \begin{subfigure}{0.48\linewidth}
        \centering
        \includegraphics[scale=0.5,trim={0 0 0 0.6cm},clip]{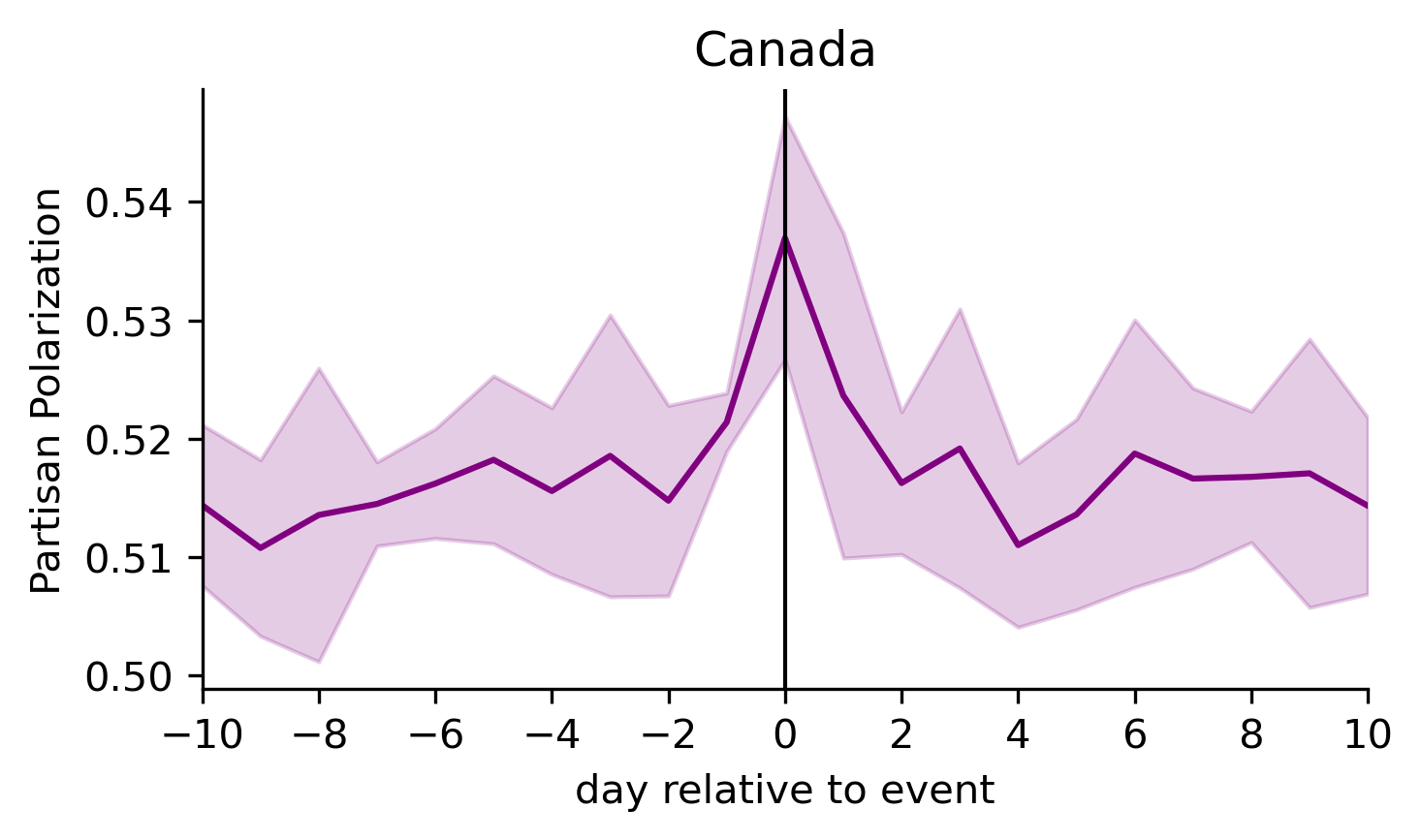}
        \caption{}
        \label{fig:cad_eventtriggered}
    \end{subfigure}

    \caption{Daily aggregated partisan polarization. For the U.S. (a, b)  and Canada (c, d), polarization is aggregated over topic by averaging over the values. We show pandemic-related new cases and deaths in background for reference. vent-triggered average polarization for identified events listed in \cref{table:peak_events}. Shaded region denotes standard deviation over the 5 events for each country.}
    \label{fig:unweighted_aggregated_poli}
\end{figure}

To complement the granular analysis presented above, we also evaluated the measure's overall responsiveness to polarizing events. In particular, we computed an average of the polarization score over topics (shown in \cref{fig:unweighted_aggregated_poli}a,b) and then performed event-triggered averaging around such events, to show how the metric varies in time before and after these particular dates on average. This aggregate result (shown in \cref{fig:unweighted_aggregated_poli}c) confirms a fast (on the order of days) and largely symmetric rise-and-decay profile around these polarization peaks.

\subsection{The Relationship between Conspiracy and Polarization}

\begin{figure}[htbp!]
 \begin{subfigure}{0.48\linewidth}
        \centering
        \includegraphics[scale=0.45]{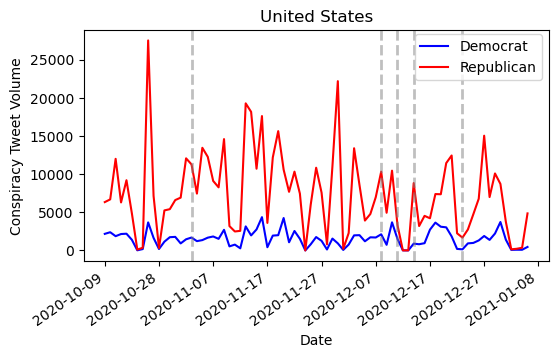}
        \caption{}
        \label{fig:us_conspiracy_vol}
    \end{subfigure}
    \hspace{0.01\linewidth}
    \begin{subfigure}{0.48\linewidth}
        \centering
        \includegraphics[scale=0.45,trim={0 0 0 .73cm},clip]{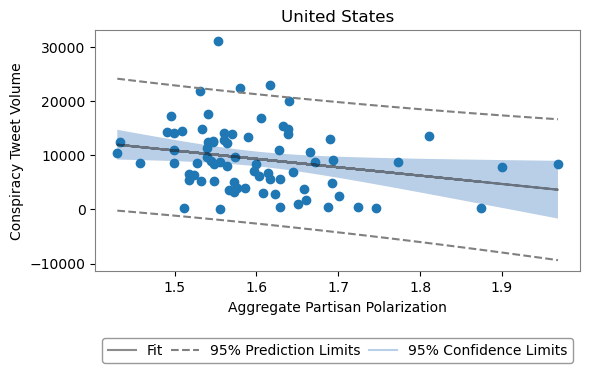}
        \caption{}
        \label{fig:us_aggregated_corr_consp}
          \end{subfigure}
    \begin{subfigure}{0.48\linewidth}
        \centering
        \includegraphics[scale=0.45]{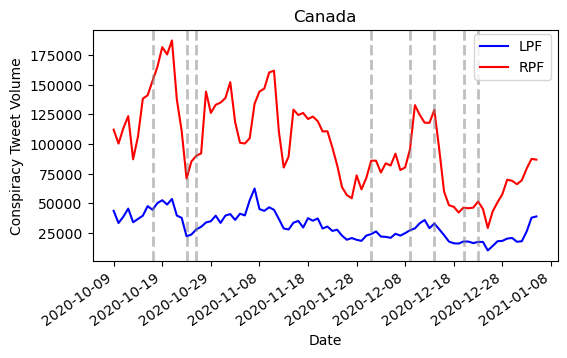}
        \caption{}
        \label{fig:cad_political_consp_vol}
    \end{subfigure}
    \hspace{0.01\linewidth}
    \begin{subfigure}{0.48\linewidth}
        \centering
        \includegraphics[scale=0.45,trim={0 0 0 .57cm},clip]{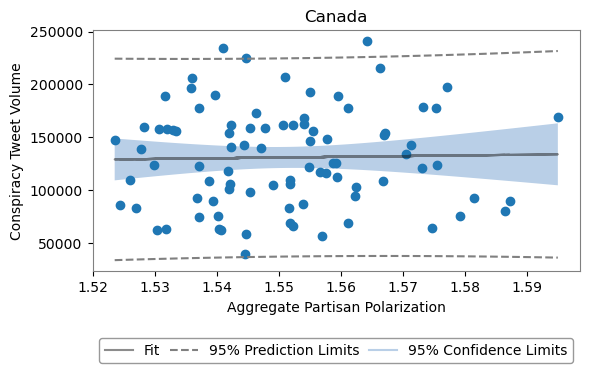}
        \caption{}
        \label{fig:cad_political_aggregated_corr_consp}
    \end{subfigure}

    \caption{Relation between the volume of conspiracy related content and the observed partisan polarization in the United States and Canada. 
    In the left column, we report the volume of conspiracy related tweets posted by users for the United States (a: affiliated with the Democrat and Republican party) and Canada (c: affiliated with the liberal (left) Party Family (LPF)---Liberal, New Democratic Party, Green, and the conservative (right) Party Family (RPF)---Conservative, People's Party for Canada). 
    On the right, we show the relation between daily partisan polarization summed over the different topics and the overall volume of conspiracy tweets. In the United States (b), we find there is a statistically significant correlation \textbf{-0.247} with CI=[-0.448,-0.023] (n = 88,  p=0.031) between these measures. In Canada (d), we find that there is no statistically significant correlation of \textbf{0.023} with CI=[-0.187,0.231] (n = 88,  p=0.831) between these measures. }
    \label{fig:aggregated_corr_polarization_and_conspiracy}
\end{figure}

Finally, we explored the relationship between conspiracy discourse and polarization using our time-resolved measurements of the number of conspiracy-related tweets. 
In particular, we compared it with the time course of the aggregate daily polarization presented in the previous section. 
The profiles broken down by progressive and liberal partisanship for U.S. and Canada are shown in \cref{fig:cad_political_consp_vol} and \cref{fig:us_conspiracy_vol}, respectively. For both countries, conservative partisans tweet conspiracy-related content in higher numbers than progressive partisans.
The correlation with aggregate daily polarization for the U.S. and Canada (\cref{fig:us_unweighted_aggregated_poli} and \cref{fig:cad_political_unweighted_aggregated_poli}) is shown in \cref{fig:us_aggregated_corr_consp} and \cref{fig:cad_political_aggregated_corr_consp}, respectively. For the U.S., 
we find a small but significant negative correlation with polarization.

\section{Discussion}\label{sec:discussion}

This article investigated regional and event-triggered variation in partisan division within social media debates surrounding the introduction of \COVID public health measures across American states and Canadian provinces. Our computational analysis was centered around quantifying partisan polarization by analyzing the language used in millions of online messages from users affiliated with different political parties. In particular, we focused on Twitter (X) discussions related to three key public health interventions during the early phases of the \COVID pandemic: lockdowns, masks, and vaccines, as well as tracking the volume of conspiracy-related tweets.
Our analysis explored the geographic heterogeneity of polarization
and identified political events that likely influenced public opinion over time.

Like several other studies before us \citep[\textit{e.g.},][]{jiang2021polarization, jiang2021social, rathje2022social,bollyky2023assessing}, we found that more right leaning states and provinces exhibited greater partisan divisions around \COVID on Twitter (X), in particular concerning topics of mask mandates and vaccine distribution.
However, we went beyond these studies to characterize the geographic heterogeneity and time course of polarization, relating features in our polarization metric to real world events. 
We looked into the relationship between polarization and public health initiatives in the U.S. and confirmed 
a strong negative correlation between partisan polarization and future vaccination rates and a moderate negative correlation between the temporal profiles of volume of conspiracy-related tweets and aggregate polarization. We did not observe similar patterns in Canada. 
 


\subsection{United States}

The early phases of the \COVID pandemic in the United States prompted a variety of public discussions online that reflected strong regional variation of partisan support \citep{jiang2020political,rao2020political,morris2021polarization,sehgal2022association,kaashoek2022evolving}. State-specific polarization obtained from our computational approach could be expected to be uniformly low, with the message content of conservative and liberal states each having internally homogeneous semantics. Instead, our analysis confirmed that polarization was notably higher in conservative states, where we found that Republican vote margins had a significant positive effect on polarization in discussions concerning masks and vaccines, after controlling for other factors.

This main result is consistent with previous studies that suggest conservative states exhibited higher levels of polarization in response to public health interventions compared to their liberal counterparts \citep{allcott2020polarization,morris2021polarization}. Nevertheless, we found that this pattern does not hold across every state. Delaware, a liberal state, exhibited a distinctly high level of polarization, likely due to the strict public health measures implemented by the governor in response to the rapid increase in the number of cases during the first wave of the pandemic \cite{Carney2020, Neiburg2020,Goldstein2021}. Similarly, low levels of polarization were observed in several conservative states, like Arkansas, Alabama, and Idaho, 
with a possible explanation coming from more unified opposition to restrictive \COVID measures like mask mandate orders \citep{adolph2022governor}. Nevertheless, most conservative states exhibited relatively high levels of polarization (\cref{fig:corr_US}). 

Our analysis did not identify a single, general cause for this relationship. That said, commonalities in each state's trajectory in the pandemic offer some clues: e.g., Mississippi, North Dakota, and Oklahoma experienced specific political decisions---such as mask mandates---made in the earlier phase of the pandemic, that led to resistance despite rising \COVID cases \citep{Wilson2020, Haines2020,rao2020political}. In Mississippi, the decision by the governor to lift the statewide mask mandate in late September could also have contributed to heightened levels of polarization \citep{Wilson2020,adolph2022governor}. A similar pattern was observed in North Dakota, South Dakota, and Oklahoma, where initial hesitancy to enforce mask mandates also appears to have led to increased partisan divisions  \cite{CarterWarren2020, German2020, ALLCOTT2020104254, Stitt2020,Macpherson2020, hallas2021variation, adolph2022governor}. This finding is not surprising, since the party affiliation of governors is the most important predictor of the widespread adoption of mask mandates \citep{mayer2022politics}. One possible explanation for our main result that can be gleaned from these anecdotes is that pandemic severity increasingly strains the more uniform opposition to restrictive health measures in more conservative states, leading them to exhibit higher levels of polarization \citep{grossman2020political,gusmano2020partisanship, adolph2022governor}. This is a distinct source of polarization than that in states with more equal distribution of partisans across competitive districts. 

Through our approach, we could also dissect how polarization varies in time over the three topics we considered: lockdowns, masks, and vaccines. These three topics exhibited similar baseline levels of polarization during the period of study, which was between the second and third waves of the pandemic, punctuated by large positive deviations that typically rise and fall quickly. The prevalence of these deviations was smallest for the lockdown topic. Its low correlation with Republican party vote share suggests that it did not act as a meaningful indicator of partisan opposition. The polarization time course for masks and vaccines, however, contained many, sharp peaks, many of which we were able to identify with a real world event. For example, South Dakota initially experienced very high levels of mask polarization, coinciding with efforts by medical authorities to promote mask-wearing, despite Governor Kristi Noem's opposition \citep{adolph2022governor, APValleyNews2020}. Her public display of opposition to Biden's suggestion of mask mandates lead to one such peak in polarization. Similarly, North Dakota also displayed early signs of increased polarization on conspiracy theories \citep{rao2020political}, which may have been exacerbated by the posthumous electoral win of a Republican candidate who died from \COVID \citep{Harmeet2020, Higgins2020, douglas2019understanding}.

The strong negative correlation between vaccines polarization and vaccination rates that we observed in the U.S. demonstrates that states with higher vaccination rates were also less polarized around this issue \citep{ye2023exploring}. The exact origin of this correlation is unclear. However, factors such as education and political ideology, which also have a strong geographic dependence, likely played a role \citep{Bollyky2023}. Indeed, higher education levels are generally associated with greater vaccine acceptance and trust in vaccine safety \cite{education_vaccine_rate}. Moreover, Democrats tend to trust the \COVID vaccines more and have been early adopters, whereas Republicans generally show lower levels of such trust \cite{latkin2021trust}.


Overall, the high levels of polarization observed in the United States relative to Canada point to a more divided society. Several studies have confirmed that conservative states and counties were less likely to adopt social distancing measures, impose mask mandates, and get vaccinated in the second and third wave of the pandemic. Our study offers new insights into these trends by  demonstrating that they correlate with regional heterogeneity in social media discourse, particularly during salient political events around health measures. We also found that this discourse reflects changes in the pandemic timeline, initially related to stay-at-home lockdown orders, followed by mask mandates, and later transitioning to vaccines as they first became available.

\subsection{Canada}
%
The Canadian set of results also suggest that partisan divisions influenced public responses to \COVID measures in this country \citep{pennycook2022beliefs}. Compared to the U.S., we found a similar, albeit much weaker association between polarization and conservative political leaning, with conservative provinces like Alberta and Saskatchewan experiencing higher levels of polarization during stricter lockdown measures than their more liberal counterparts, such as British Columbia and Ontario \citep{cheung2021bank, Rowe2020}. Polarization levels varied over smaller and medium-sized provinces as well, measured as relatively high for New Brunswick and low for Nunavut, where the rates of \COVID infections remained relatively low during the pandemic (the sole COVID-free jurisdiction in North America until November 2020) \citep{cameron2021variation,Akanteva2023}.
Additionally, we also found that polarization surrounding mask mandates and vaccines were not homogeneously distributed across provinces. 

Among the Canadian provinces, Quebec is an interesting case for our analysis of polarization. For example, our results confirmed that Quebec had the highest level of partisan division over vaccines, but also the highest reported incidence of \COVID in Canada during the first and second waves of the pandemic \citep{dube2021covid}. Quebec's unique approach to managing the pandemic with its more restrictive measures relative to other provinces is also somewhat reflected in our results. After a relative hiatus with several restrictions relaxed in the summer of 2020, Quebec once again became the epicenter of the pandemic in the fall \citep{shim2021regional}. This resurgence led to the reinstatement of strict pandemic control measures and a ban on public demonstrations following significant anti-government protests against lockdowns and mask mandates \citep{Montpetit2020,Gazette2020}. These events also coincided with an increase in online conflicts, promoted by Canadian far-right populist rhetoric and conspiracy theories on Twitter (X) \citep{chaput2021figures}. It is important to note, however, that most of these conversations in Canada were heavily influenced by discussions in the US, with Canadians retweeting American vaccine-related content 8 times as often as Canadian content during the period covered by our study \citep{Owen2020,boucher2021analyzing}. Likewise, vaccine hesitancy was also linked to political affiliation in Canada, with those supporting the Conservative Party more likely to refuse vaccination \citep{burns2024examining}. 




As in the U.S. case, the polarization time course computed for Canada also exhibits spikes observed around key events like protests against lockdown measures, mask mandates, and vaccine roll-outs. These findings suggest that public reactions to significant political and social events during the pandemic are reflected in the measure of polarization we use. We did not observe the negative correlation between polarization and volume of conspiracy-related tweets that we saw in the U.S. case. This contrasts with \citep{Owen2020}, who found a reduction in negative sentiment in Canadian vaccine-related tweets between January and December 2020. 
The relationship between polarization and sentiment is complex and long-term trends are likely driven by processes besides pure volume of discussion around conspiracy theories \citep{van2021social}.

Finally, the relatively lower influence of polarization on vaccine attitudes may be attributed to the country's more widespread vaccine mandates \citep{karaivanov2022covid,cameron-blake2021a}. This prevalence, along with higher levels of trust in politicians \citep{mansoor2021citizens} and social capital \citep{hetherington2017political, makridis2021social}, could have contributed to a broader acceptance of \COVID health interventions  \cite{boucher2021analyzing,burns2024examining}. Indeed, there was a rare `cross-partisan consensus' among Canadians regarding emergency measures in the early stages of the pandemic \citep{Merkley2020-pg}. This consensus, however, was not mirrored on social media, where conspiracy theories widely circulated \citep{Bridgman2020, boucher2021analyzing}. Overall, our results indicate that online discussions surrounding lockdowns, masks, and vaccines did mirror polarization, and were shaped by regional reactions to events and circumstances specific to Canadian provinces.


\subsection{Limitations}

While our method offers valuable insights, it comes with certain limitations. First, we viewed partisan polarization only through the proxy of semantic similarity. This choice may in certain cases obscure some signals not captured by the semantic embedding representation. Second, specifically in the Canadian context, we categorized users into liberal (left) and conservative (right) party family groups. During the manual annotation of Twitter (X) profiles, we encountered few users who identified as supporters of the Bloc Quebecois political party; therefore, we opted to exclude them from the analysis. Additionally, our classification of users into liberal and conservative partisan groups is based on self-reported information, which may not be entirely accurate. Third, it is important to note that our analysis is based on Twitter (X) data, which may not fully capture the views and sentiments of the broader American and Canadian public. Fourth, our analysis is restricted to tweets in English. In the context of Canada, this means we are capturing only or primarily the perspectives of either anglophones or bilingual francophones, which could potentially bias our data; for example, the high levels of polarization observed in Quebec on \COVID measures may be influenced by this language bias. Finally, while several of our analyses rely on correlations, it is crucial to remember that these results do not imply causation; the relationship between polarization and public health measures is complex and multi-dimensional.

\subsection{Conclusion}

To conclude, our method has provided valuable insights into the dynamics of partisan polarization during the \COVID pandemic. Political ideology, public trust, and key events have emerged as important factors influencing public discussions on pandemic-related issues in the United States and Canada. By combining our polarization measure with other data, researchers and practitioners can better understand how polarization varies across location, time, and specific issues. This knowledge could help in detecting particularly polarizing discussions on social media and in developing communication strategies to mitigate the spread of misinformation, both for the current pandemic and for future health-related crises.

The differences observed between these two countries are somewhat harder to explain. Our analysis, along with insights from recent studies, suggests that Canadian responses to public health measures could explain the lower levels of polarization found in Canada. Indeed, there was a significant consensus on the effectiveness of stay-at-home orders (i.e., lockdowns), mask mandates, and vaccines not only at the federal and provincial levels, but also within the news media. And unlike the U.S., where an important number of Republican leaders aligned with Trump's anti-mask and anti-lockdown positions, the pandemic did not become a salient partisan issue within a political campaign until much later in 2021. Prior to this, the opposition to public health measures in Canada was primarily found in online communities, outside of the mainstream media and political parties, where protesters remained heavily influenced by American sources. Although our results suggest that social networks contributed to the diffusion of these opinions during the \COVID pandemic, more work needs to be done to quantify the impact of online communities interactions on polarization.

\section{Methodology}\label{sec:methodology}
In the following section, we describe in detail our text-based measurement of partisan polarization. We first explain the data collection process. We then show how we classified tweets into respective topics, geo-located users and grouped them by party affiliation. Finally, we describe the equation used to measure partisan polarization as well as our approximation algorithm. Figure \ref{fig:Methodology_Flow} provides a visual overview of our process in measuring partisan polarization. For additional details, please refer to Section \ref{sec:methodology_cont} in the Supplementary Material.

\subsection{Data Collection}

\subsubsection{Twitter (X) Data}
We used Twitter's (X) official API to collect 1\% of real-time tweets for Canada and the United States from October \nth{9}, 2020 to January \nth{4}, 2021. This represents 231,841,790 tweets and 4,765,115 users for Canada (a dataset filtered for COVID and politics)  and 387,090,097 tweets and 23,758,112 users for the United States (a dataset filtered for election politics). We fed the following list of keywords in the API to filter relevant tweets:

{\noindent \textbf{Canada}}: {\scriptsize `trudeau', `legault', `doug ford', `pallister', `horgan', `scott moe', `jason kenney', `dwight ball', `blaine higgs', `stephan mcneil', `cdnpoli', `canpol', `cdnmedia', `mcga', `covidcanada' and all combinations of `covid' or `coronavirus' as the prefix and the (full \& abbreviated) name of each provinces and territories as the suffix}. 

{\noindent \textbf{United States}}: {\scriptsize `JoeBiden', `DonaldTrump', `Biden', `Trump', `vote', `election', `2020Elections', `Elections2020', `PresidentElectJoe', `MAGA', `BidenHaris2020', `Election2020'}.






\subsection{\COVID Vaccination Rate}
Similar to the \COVID pandemic data, we also used the officially reported vaccination rate of the populations. We used the vaccination rates one year later compared to the Twitter (X) data, as \COVID vaccines were created and approved at the very end of our data collection process. Thus, the vaccination rates are for those who obtained at least two doses. For Canada, this is the $numtotal\_fully$ from the government's \href{https://health-infobase.canada.ca/covid-19/vaccination-coverage/}{vaccine coverage} map. We normalize this column by Canada's 2021 population per province or territory. For the United States, we use the $people\_fully\_vaccinated\_per\_hundred$ reported in the \href{https://covid.cdc.gov/covid-data-tracker/#vaccinations_vacc-people-booster-percent-pop5}{COVID Data Tracker} from the CDC. 

\subsection{Classifying Tweets By Topics}

For this study, we looked into three key topics for \COVID: lockdowns, masks and vaccines. We also looked into conspiracy theories. For each topic, tweets were classified as \textit{relevant} or \textit{irrelevant} to the topic based on whether they contained at least one of the topic-specific keywords. For conspiracy-related tweets, \textit{relevant} means that the content is related to \COVID conspiracy theories (either supporting or opposing). A tweet can belong to more than one topic.

We first used a hashtag-based filtering step. We extracted all hashtags within our dataset, ordered it by frequency, and discarded those that appeared less than 100 times. This filtered list contained 3,600 hashtags for Canada and 18,000 for the United States. Two political scientists manually annotated this list with topic and relevance labels. The list was narrowed to only those hashtags labeled as relevant, resulting in 631 relevant hashtags. We then merged these with hashtags identified in previous studies for the same topics---i.e., from refs. \cite{Kouzy2020,Al-Ramahi,Ahmed_2020}. 

For Canada, this process resulted in 46,636,206 tweets and 1,757,675 users that shared content related to \COVID. For the United States, this represents 12,552,213 tweets and 2,657,355 users. Using the RoBERTa-base model \cite{liu2019roberta} from HuggingFace, we further pre-trained this model on the respective \COVID tweets from each country dataset---i.e., performing a self-supervised learning on predicting masked words within tweets. This results in two different country-specific pre-trained language models for \COVID tweets.

We then randomly sampled 200 \textit{relevant} and 200 \textit{irrelevant} tweets per topic from each dataset, for a total of 1,600 tweets. The same two political scientists manually reviewed each tweet separately to determine if the tweet was relevant/irrelevant. We discarded tweets where the annotators could not reach a consensus. We then trained the respective pre-trained RoBERTa-base model on each dataset to classify by topics---i.e., 4 topic language models per country, for a total of 8 language models. We report the support, Cohen Kappa, F1-score and number of tweets we extracted for each topic within each dataset in Table \ref{table:Tweet_Classification_Statistics}. Our analysis achieved a near perfect F1-score for each of these topics.

\subsection{Classifying Users by Geo-Location}
We wanted to quantify users in each province or state represented in our data, as the users retrieved from Twitter (X) could be imbalanced relative to region population size. For this, we geolocated all users with an explicit location provided in the location field, a free-form text, as part of their profile information. We process the information with \href{https://www.openstreetmap.org}{Open Street Map} and the \href{https://developers.arcgis.com/python}{ArcGIS API}. Both of these return a latitude and longitude if a location was found. We set the threshold for a clear geolocation if both API responded with a latitude and longitude that was within one degree of each other. We found this to be more accurate, and preferable to using a pre-trained Named Entity Recognition algorithm; most of the user-provided locations can be handled by these Geographic Information System APIs, and the APIs could also provide important details such as the country, state and city. 
We correlated the geolocated users with the official population census for each country's region.  In total, we geolocated \textbf{282,454} users with strong correlation of \textbf{0.92} (n=13, p=6.20e-06, CI=[0.76, 0.98]) for Canada and \textbf{757,601} users with strong correlation of \textbf{0.98} (n=52, p=9.27e-35, CI=[0.96, 0.99]) for the United States. This means that each region is well represented in our data. Further information is reported in Table \ref{table:Geolocation_correlation}.

\subsection{Classifying Users By Party Affiliation}

We determine a user's party affiliation using a two step approach. First, we classify politically explicit users based on their profiles. We then use the predictions from this profile classifier as labels to train a classifier based on the user activity. We report the support, F1-score, and number of users we classified for each party within each dataset in Table \ref{table:CAD_UPA_Classification_Accuracy} for Canada and Table \ref{table:US_UPA_Classification_Accuracy} for the United States. We achieve a macro-F1-score of 91\% for both Canada and the United States. 

\paragraph{Profile Classifier}

As a preprocessing step, we filter out users that are not politically explicit. \textit{Politically explicit users} are those whose profile description contains at least one political keyword defined for any political party. For Canada, we focused on the five main political parties: Conservative, Green, Liberal, New Democratic Party and People's Party. For the United States, we focused on the Democratic and Republican parties. The following is the set of keywords we have per party:

{\noindent \textbf{Canada}}:

\textit{Conservative} - {\scriptsize `erin o'toole', `andrew scheer', `conservative', `conservative party', `cpc', `cpc2021', `cpc2019', `conservative party of canada'}

\textit{Green} - {\scriptsize `annamie paul', `green party', `gpc', `gpc2019', `gpc2021', `green party of canada'}

{\textit{Liberal}} - {\scriptsize `justin trudeau', `liberal', `liberal party', `lpc', `lpc2021', `lpc2021', `lpc2019', `liberal party of canada'}

{\textit{New Democratic Party}} - {\scriptsize `jagmeeet singh', `new democrat', `new democrats', `new democratic party', `ndp', `ndp2021', `ndp2019'}

{\textit{People's Party}} - {\scriptsize `maxime bernier', `people's party', `ppc', `ppc2019', `ppc2021', `people's party of canada'}

{\noindent \textbf{United States}}:

{\textit{Democrat}} - {\scriptsize `liberal,' `progressive,' `democrat,' `biden'}

{\textit{Republican}} - {\scriptsize `conservative,' `gop,' `republican,' `trump,'}

We then randomly selected a set of \textit{politically explicit} users for each party to be manually annotated by two political scientists. We only use party labels that both annotators agreed upon. The Cohen Kappa score for the pair of annotation sets is \textbf{0.74} and \textbf{0.76} for Canada and the United States respectively. We then train a RoBERTA-large model with a 80-20 train-test split to determine user party affiliation (see the profile classifier section for more details). Exact numbers can be found in their respective tables in the supplementary.

\paragraph{Activity Classifier}

For this second phase, we make use of the respective RoBERTa-base model pre-trained on \COVID tweets for each dataset to extract the tweet embeddings (768-dimensional vector). We then generate user embeddings (768-dimensional vector) by pooling together (mean aggregation) all tweet embeddings from that user.

We then train an MLP consisting of two fully connected layers with the user embeddings as input to predict the party affiliation. Before training, we filtered out users based on their activity  $\alpha$ (i.e., number of \COVID-related tweets in the dataset). 
We performed a hyperparameter search for $\alpha$ among $\{1, 3, 5, 10, 15, 20\}$ using 5-fold cross validation. This was 5 tweets for Canada and 10 tweets for the United States.

Specifically for Canada, we found that the MLP could not distinguish the parties sufficiently. Hence, we grouped the parties based on their partisan leaning. The liberal (left) party family included the Liberal Party, New Democratic Party and Green Party while the conservative (right) party family included the Conservative Party and People's Party. We removed supporters of other minor parties and the Bloc Quebecois.

\paragraph{External Evaluation}

Following the best practice for evaluating party affiliation predictions \cite{barbera2015birds}, we matched Twitter (X) users from the United States with the primary voter registration records available for five states: Ohio, New York, Florida, Arkansas, North Carolina, as well as Washington DC. We describe this procedure and its result in more details in the supplementary material \ref{table:voter_registration_stats}. We achieve an accuracy of 74.35\% for the profile classifier and 73.35\% for the activity classifier. Both classifiers are binary, but users can also be independent or support a third party, despite an ideology (and behavior/voting) that aligns strongly with one of the two main parties. They can also have an outdated registration that no longer reflects the beliefs they currently hold and express on Twitter (X). Therefore, although accuracy here is lower than in our training model, it still indicates our classification is acceptable and on par with the standard for this type of evaluation \citep{barbera2015birds}.

\subsection{Measuring Partisan Polarization}

Given a set of political parties $\mathcal{P}$ and a set of given user embeddings $\mathcal{U}=\{{u^{(1)}, u^{(2)}, \dots,  u^{(n)}}\}$ where $u^{(i)} \in \mathbb{R}^{768}$ and $n$ is the number of users, we measure polarization as follows.

We first look at the distance and dispersion between each party, $p\in \mathcal{P}$. We base our measure on the C-index of Hubert \cite{Hubert1976AGS} to quantify the extent of clustering and overlap of each political party. This is done by first calculating the sum of inter-cluster distances:

\begin{equation}
S_{w} = \frac{1}{2}\sum_{p \in \mathcal{P}} \sum_{u,v \in p} ||u-v|| 
\end{equation}
We then normalize this value based on its minimum and maximum possible ranges, $S_{min}$ and $S_{max}$. These correspond to the sum of the $m$ smallest (resp. largest) distances between points in $\mathcal{U}$; where $m =  \sum_{p \in \mathcal{P}} \frac{|p|(|p|-1)}{2}$. 

Based on these, we define our polarization index $\poli$ as:  
\begin{equation} \label{eq:C_Index} 
\poli = \frac{S_{max} - S_{w}}{S_{max} - S_{min}}
\end{equation}

The minimum value 0 represents no polarization, whereas the maximum value 1 represents the most extreme possible polarization, i.e., $p \in \mathcal{P}$ are completely isolated from each other. 

\subsection{Approximation of Polarization}

Equation \ref{eq:C_Index} is not scalable to large $n$, as it is $O(n^2 \log(n^2))$. We approximate it by sub-sampling a sufficient set of users which enables us to scale to a large number of users. To determine the minimum sample size needed, we use the coefficient of variation, which is defined as $\frac{std}{mean}$ and expressed as a percentage. Generally, a coefficient of variation under 10 gives reasonable results \citep{coefficientOfVariationUse}.

Algorithm \ref{alg:poli} summarizes this procedure. This approximation allows us to scale our measure significantly without compromising accuracy. One loop of the approximation has a time complexity of $O(rf^2 * n^2 \log((n)^2))$ where $r$ is the repeat count and $f$ is the fraction (e.g., 0.01). From our testing, we know that at large values of $n$, the fraction needed rarely increases, so only one loop is required. Therefore, the time saved is $rf^{2}$ for large values of $n$.

We test the accuracy of approximation in Algorithm 1 over the daily lockdown and vaccine tweets. In Figure \ref{fig:C_Index_Quality}, we plot the total number of users against the absolute error (and its standard deviation) of the approximated $poli$ compared to the exact value, binned for every 1,000 users. We observe a dramatic drop of the absolute error term at around 3,000 users. When we reach the 10,000 users value, the absolute error is usually below 0.001. In Figure \ref{fig:C_Index_Time_Saved}, we plot the total number of users against the time saved in running the approximation algorithm compared to running the exact $poli$, binned also for every 1,000 users. We observe that the time saved is exponential to the number of users. We note that at around 50,000 users, the approximation rarely needs to increase the fraction of users sampled.  

These findings confirm that we can accurately approximate $poli$ for large-scale data that is impossible to measure exactly because of memory constraints. As $poli$ relies heavily on finding pairwise distances (time and memory intensive), we see from our analysis that a sampling approach can save both time and memory exponentially.

\begin{algorithm}
\caption{Approximating $\poli$}\label{alg:poli}
\begin{algorithmic}[1]
\Require $\mathcal{U}$, $fraction$, $epsilon$, $step\_size$, $repeats$
\While{ $cv\_poli > epsilon$ }
    \State $poli\_indices \gets []$
    \State $num\_of\_runs \gets 0 $

    \For{$i$ in $repeats$}
        \State $\mathcal{U}_s \gets sample(\mathcal{U}, fraction)$
        \State $poli\_indices.append(poli(\mathcal{U}_s))$ 
    \EndFor

    \State $mean\_poli \gets mean(poli\_indices)$ 
    \State $std\_poli \gets std(poli\_indices)$ 
    \State $cv\_poli \gets std\_poli / mean\_poli$ 

    \State $fraction \gets fraction + step\_size$
\EndWhile
\State \textbf{return} $mean\_poli$
\end{algorithmic}
\end{algorithm}

\section{Acknowledgements}\label{sec:acknowledgments}
This research is supported by CIFAR AI Catalyst Grants and Canada CIFAR AI Research Chair funding.

\bibliography{ref}

\section{Extended Results}

\subsection{Regional Variations of Partisan Polarization}
\begin{figure}[ht!]
    \centering
    \includegraphics[scale=0.65]{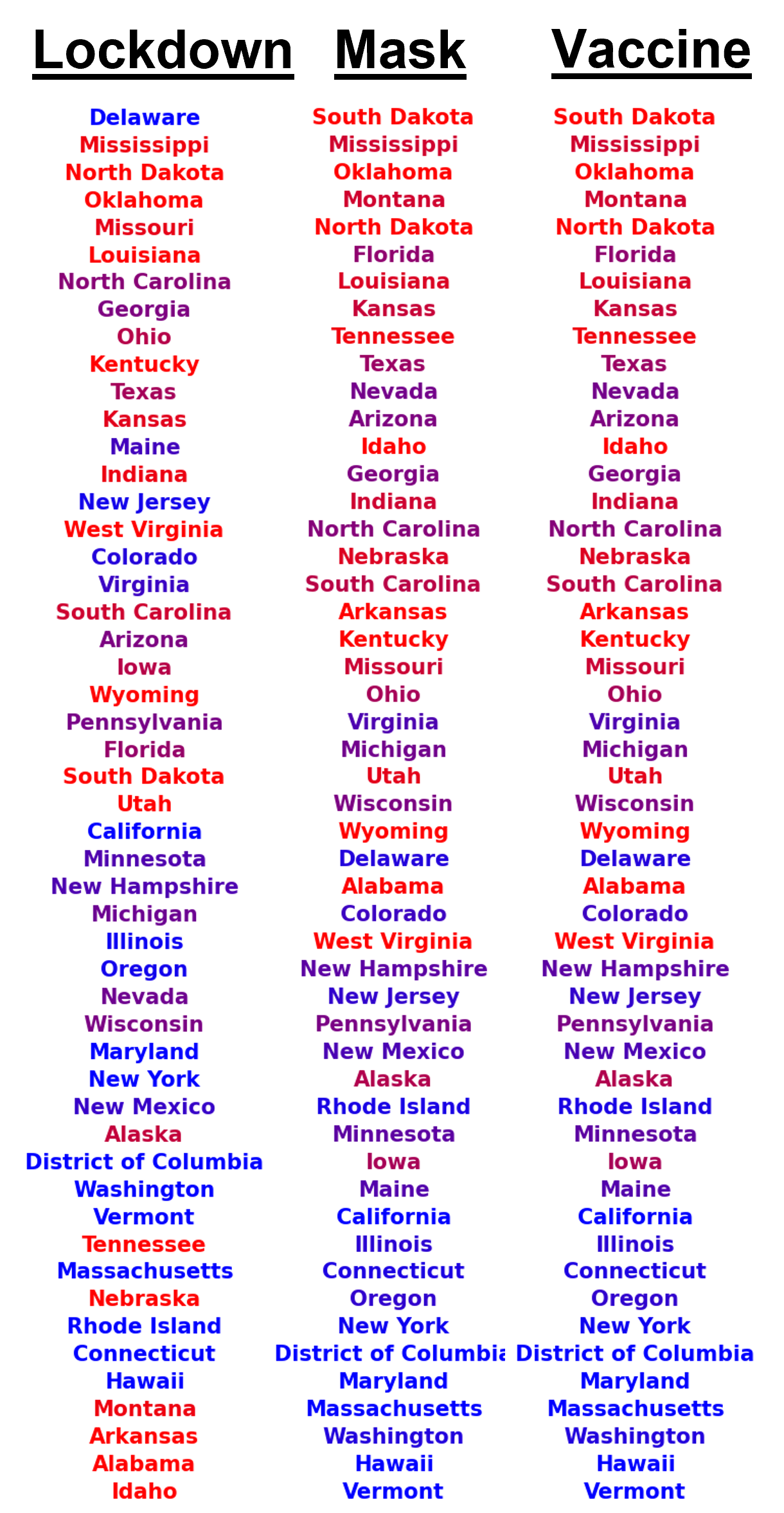}
    \caption{Ranking of American states by partisan polarization per topic. Ranking of 1 signifies the highest average weekly polarization between October \nth{11}, 2020 to January \nth{3}, 2021 (12 weeks). State names are colored based on the party ratio from the 2020 United States Presidential Election, where more blue means more users voted for the Democratic Party and more red means more users voted the Republican Party. We can see that red states are mostly ranked higher than blue states.}
    \label{fig:us_state_total_weekly_pol}
\end{figure}

\begin{figure}[ht!]
    \centering
    \includegraphics[scale=0.42]{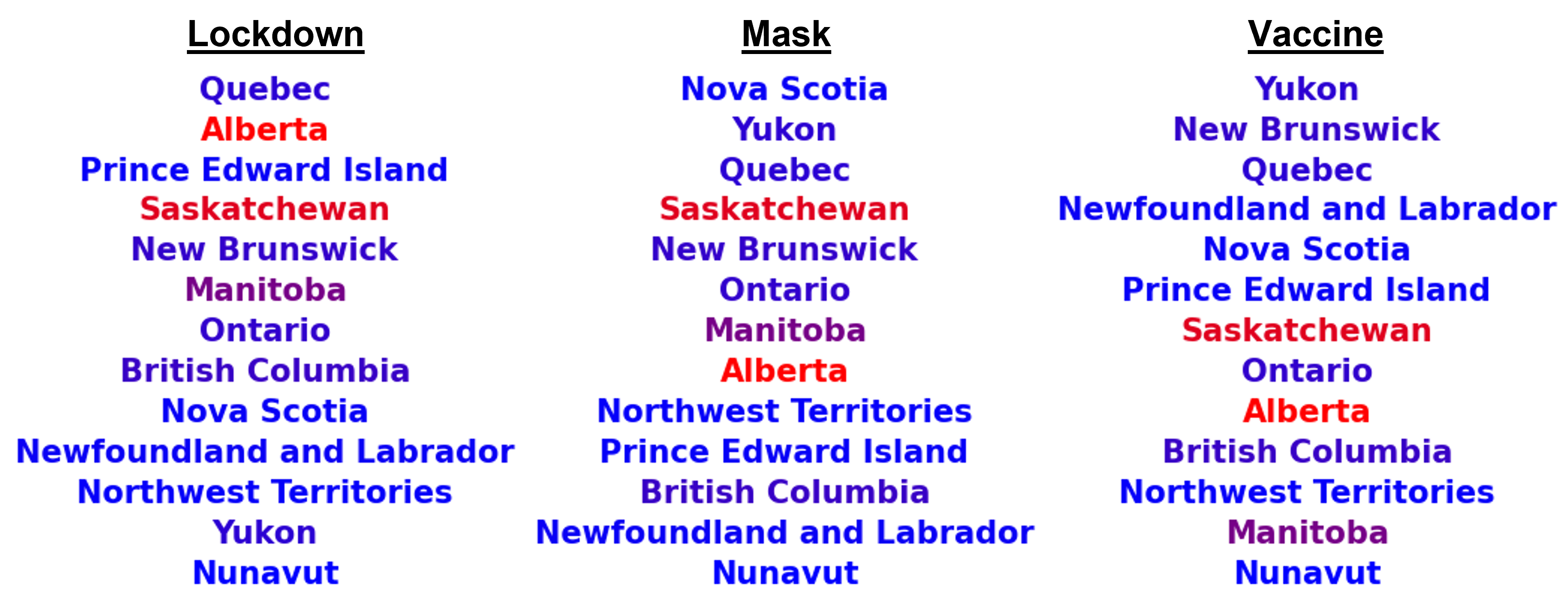}
    \caption{Partisan polarization ranking of Canadian provinces and territories per topic. A ranking of 1 signifies the highest average weekly polarization between October \nth{11}, 2020 to January \nth{3}, 2021 (12 weeks). Province or territory names are colored based on the party ratio from Canada's 2021 Federal Election, where more blue means more users from the liberal (left) party family (Liberal, New Democratic Party, Green), and more red means more users from the conservative (right) party family (Conservative, People's Party).}
    \label{fig:cad_state_total_weekly_pol}
\end{figure}

In \cref{fig:us_state_total_weekly_pol} and \cref{fig:cad_state_total_weekly_pol}, we show the ranking of the regions (American states and Canadian provinces and territories, respectively) by partisan polarization per topic.

 \begin{figure}
     \centering
     \includegraphics[width=\linewidth]{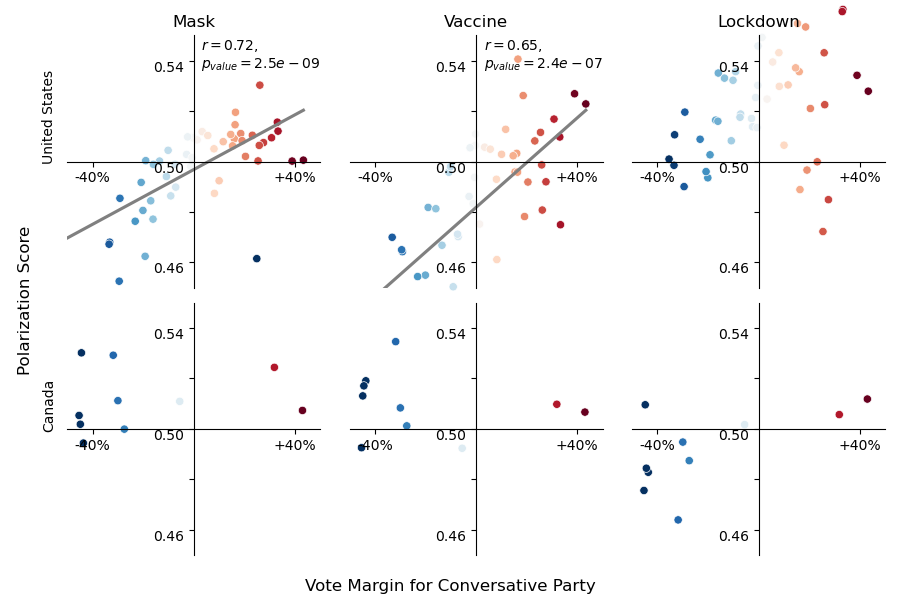}
     \caption{Correlation between Polarization Score and Vote Margin for Conservative Party. Colors are vote margin. Respective Pearson $r$ correlation and corresponding $p$-value is shown in each panel. Significant correlation between Polarization Score and Vote Margin is found for the US discourse on Masks and on Vaccines.}
     \label{fig:corr_US_1}
 \end{figure}

In \cref{fig:corr_US_1}, we show the correlation in the polarization scores and relative vote margins. The latter shows that states in which the margin by which Republican party votes exceeded those of the Democractic party correlates significantly with the amount of polarization exhibited by the Twitter (X) discourse in those states, when conditioning the discourse on masks and on vaccines, but though significantly when conditioning on lockdowns.

\subsection{The Relationship between Regional Vaccine Polarization and Vaccination Rates in Canada}

\begin{figure}[h!]
    \centering
    \includegraphics[scale=0.5]{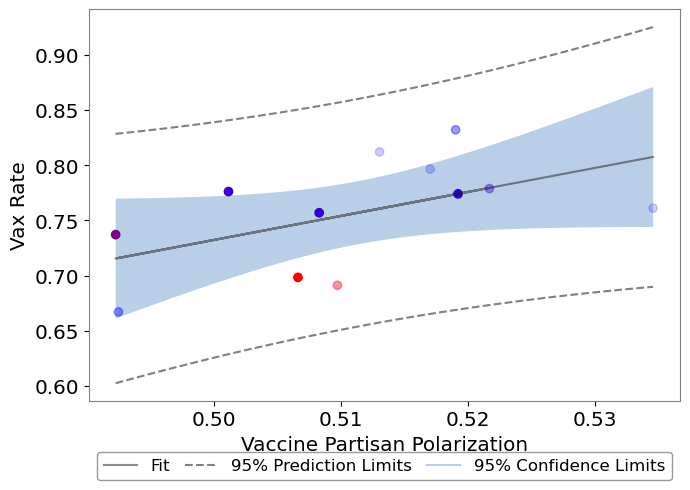}
    \caption{Relation between vaccine polarization and vaccination rates in Canada. The correlation is \textbf{0.74} with CI = [0.31, 0.92] (n = 12, p = 0.004). The Vaccine Partisan Polarization for each province or territory is computed weekly and averaged over 12 weeks from October \nth{11}, 2020 to January \nth{3}, 2021. Official vaccination rates for different regions are obtained from Statistic Canada. The Vaccination Rate is also averaged weekly for the similar period of time a year into future to be after the vaccines were rolled out, i.e. October \nth{11}, 2021 to January \nth{3}, 2022. Color for the scatter plots is determined by the respective party ratio from the 2021 Canadian federal election.}
    \label{fig:CAD_AVG_VAX_POL_v_VAX_RATE}
\end{figure}

In Figure \ref{fig:CAD_AVG_VAX_POL_v_VAX_RATE}, we remove Nunavut as an outlier because of its very small population. While we get strong positive correlation with vaccination rate, it is over a relatively low number of points. In Canada, vaccines were mandated, requiring vaccine passports to be served in public areas. We assume that vaccine partisan polarization increases, as people are not happy with being forced to be vaccinated, but most of the population still are vaccinated. However, with the few points, we do not have a definite conclusion for this result. 

\subsection{Specific Regional Partisan Polarization and COVID-19 Deaths}

\begin{figure}[ht!]
    \begin{subfigure}{0.48\linewidth}
        \centering
        \includegraphics[scale=0.43]{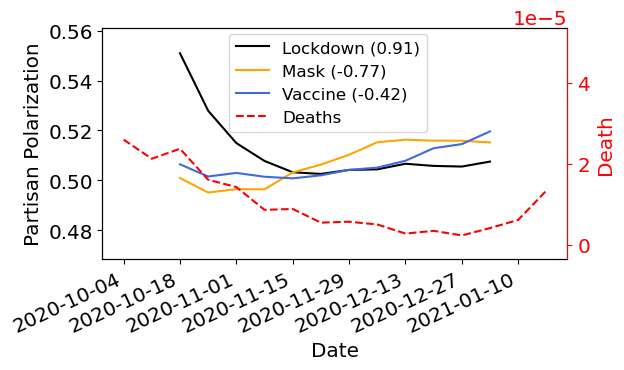}
        \caption{Weekly Partisan Polarization \& Death Rate in \textbf{Alberta}}
        \label{fig:CAD_ALBERTA_POL_DEATHS}
    \end{subfigure}
    \hspace{0.02\linewidth}
    \begin{subfigure}{0.48\linewidth}
        \centering
        \includegraphics[scale=0.43]{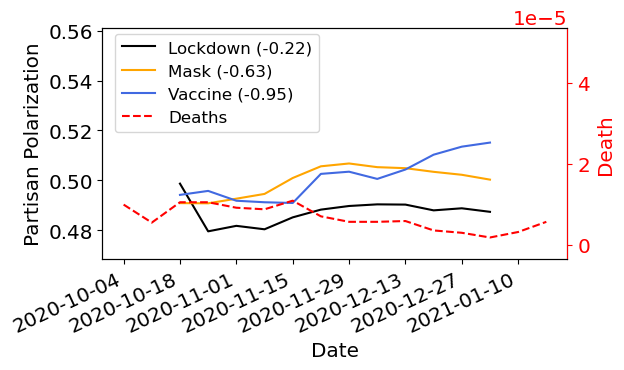}
        \caption{Weekly Partisan Polarization \& Death Rate in \textbf{British Columbia}}
        \label{fig:CAD_BRITISH_COLUMBIA_POL_DEATHS}
    \end{subfigure}
    \\   
    \begin{subfigure}{0.48\linewidth}
        \centering
        \includegraphics[scale=0.43]{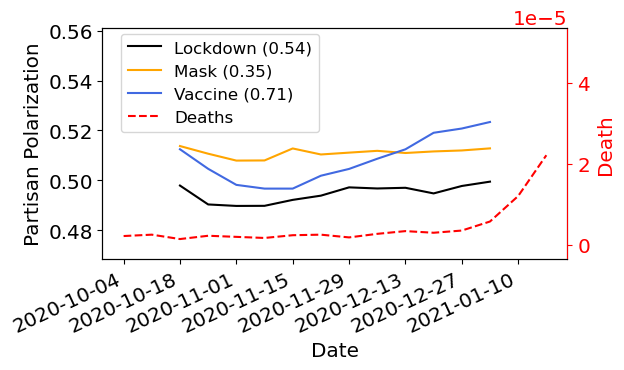}
        \caption{Weekly Partisan Polarization \& Death Rate in \textbf{Ontario}}
        \label{fig:CAD_ONTARIO_POL_DEATHS}
    \end{subfigure}
    \hspace{0.02\linewidth}
    \begin{subfigure}{0.48\linewidth}
        \centering
        \includegraphics[scale=0.43]{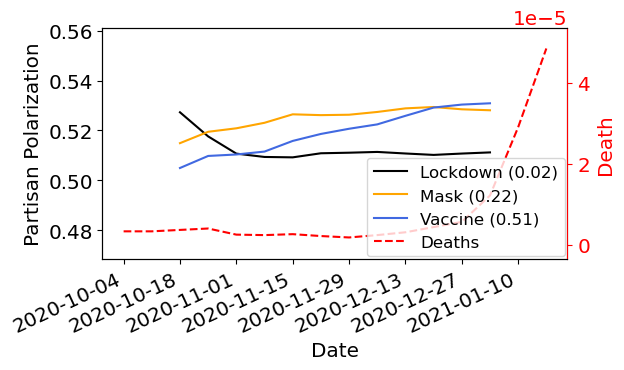}
        \caption{Weekly Partisan Polarization \& Death Rate in \textbf{Quebec}}
        \label{fig:CAD_QUEBEC_POL_DEATHS}
    \end{subfigure}
    \caption{Weekly trends of partisan polarization in \textbf{Canada} for the top 4 largest provinces from October \nth{11}, 2020 to January \nth{3}, 2021. We report the average death rate (red dotted line) per week and report the correlation between the topic-specific correlation with the death rate in the brackets in the legend.}
    \label{fig:CAD_PROV_POL_DEATHS}
\end{figure}

\begin{figure}[ht!]
    \begin{subfigure}{0.48\linewidth}
        \centering
        \includegraphics[scale=0.43]{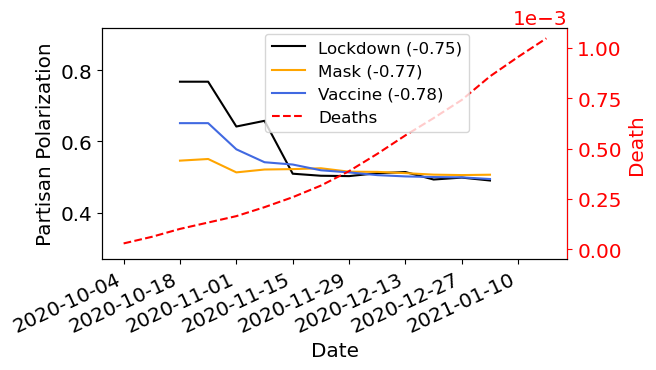}
        \caption{Weekly Partisan Polarization \& Death Rate in \textbf{Mississippi}}
        \label{fig:US_MISSISSIPPI_POL_DEATHS}
    \end{subfigure}
    \hspace{0.02\linewidth}
    \begin{subfigure}{0.48\linewidth}
        \centering
        \includegraphics[scale=0.43]{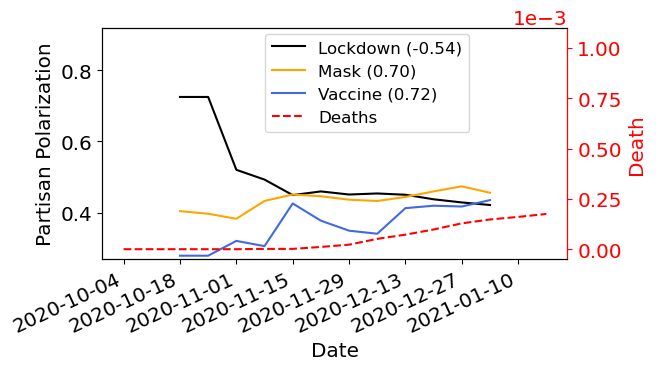}
        \caption{Weekly Partisan Polarization \& Death Rate in \textbf{Vermont}}
        \label{fig:US_VERMONT_POL_DEATHS}
    \end{subfigure}
    \\
    \begin{subfigure}{0.48\linewidth}
        \centering
        \includegraphics[scale=0.43]{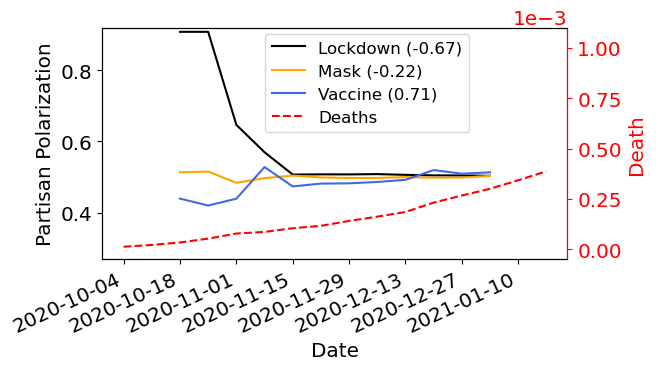}
        \caption{Weekly Partisan Polarization \& Death Rate in \textbf{Delaware}}
        \label{fig:US_DELWARE_POL_DEATHS}
    \end{subfigure}
    \hspace{0.02\linewidth}
    \begin{subfigure}{0.48\linewidth}
        \centering
        \includegraphics[scale=0.43]{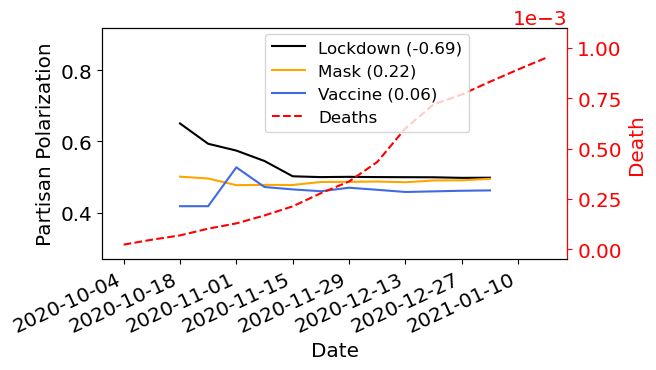}
        \caption{Weekly Partisan Polarization \& Death Rate in \textbf{Iowa}}
        \label{fig:US_IOWA_POL_DEATHS}
    \end{subfigure}
    \\
    \caption{Weekly trends of partisan polarization in the \textbf{United States} for highest ranked state overall (a), lowest ranked state (b), highest ranked liberal state (c) and lowest ranked conservative state (d) from October \nth{11}, 2020 to January \nth{3}, 2021. We report the average death rate (red dotted line) per week and report the correlation between the topic-specific correlation with the death rate in the brackets in the legend.}
    \label{fig:US_STATE_POL_DEATHS}
\end{figure}

Here, we investigate the topic-specific polarization over time and how it relates to the reported number of Deaths for \COVID for specific regions in Figure \ref{fig:CAD_PROV_POL_DEATHS} for Canada and in Figure \ref{fig:US_STATE_POL_DEATHS} for the United States.

\subsection{Correlation Matrices}

\begin{table}[ht!]
\captionsetup{width=\linewidth}
\caption{Correlation Matrix between Topic Polarization and External Data in Canada. Bolded means $p < 0.001$. Italicized means $p < 0.01$. Underline means $p < 0.05$. Background color of green or red signifies the positive or negative correlation for significant p-values only.}
\label{table:cad_political_corr_matrix}
\begin{tabular}{ll|llll}
                             &                                                                 & Cases                                                                                                                        & Deaths                                                                                                             & Conspiracy (Volume)                                                                                                & Stringency Index                                                                                                   \\ \hline
                             & Polarization                                                    & \cellcolor[HTML]{FD6864}\textit{\textbf{\begin{tabular}[c]{@{}l@{}}-0.338 \\ CI={[}-0.511,-0.138{]}\\ p=0.001\end{tabular}}} & \cellcolor[HTML]{FD6864}\textit{\begin{tabular}[c]{@{}l@{}}-0.280\\ CI={[}-0.462,-0.075{]}\\ p=0.008\end{tabular}} & \begin{tabular}[c]{@{}l@{}}0.190\\ CI={[}-0.020,0.384{]}\\ p=0.076\end{tabular}                                    & \cellcolor[HTML]{FD6864}\textbf{\begin{tabular}[c]{@{}l@{}}-0.518\\ CI={[}-0.657,-0.347{]}\\ p=0.000\end{tabular}} \\
                             & Volume                                                          & \cellcolor[HTML]{32CB00}\begin{tabular}[c]{@{}l@{}}0.244\\ CI={[}0.037,0.432{]}\\ p=0.022\end{tabular}                       & \begin{tabular}[c]{@{}l@{}}0.171\\ CI={[}-0.040,0.367{]}\\ p=0.112\end{tabular}                                    & \begin{tabular}[c]{@{}l@{}}-0.084\\ CI={[}-0.288,0.128{]}\\ p=0.437\end{tabular}                                   & \cellcolor[HTML]{32CB00}\begin{tabular}[c]{@{}l@{}}0.259\\ CI={[}0.052,0.444{]}\\ p=0.015\end{tabular}             \\
                             & \% Volume                                                       & \cellcolor[HTML]{FD6864}\textbf{\begin{tabular}[c]{@{}l@{}}-0.368\\ CI={[}-0.536,-0.172{]}\\ p=0.000\end{tabular}}           & \cellcolor[HTML]{FD6864}\textit{\begin{tabular}[c]{@{}l@{}}-0.340\\ CI={[}-0.513,-0.141{]}\\ p=0.001\end{tabular}} & \begin{tabular}[c]{@{}l@{}}0.149\\ CI={[}-0.062,0.348{]}\\ p=0.165\end{tabular}                                    & \cellcolor[HTML]{FD6864}\textbf{\begin{tabular}[c]{@{}l@{}}-0.391\\ CI={[}-0.555,-0.198{]}\\ p=0.000\end{tabular}} \\
\multirow{-4}{*}{Lockdown}   & \begin{tabular}[c]{@{}l@{}}Weighted\\ Polarization\end{tabular} & \cellcolor[HTML]{FD6864}\textbf{\begin{tabular}[c]{@{}l@{}}-0.402\\ CI={[}-0.564,-0.210{]}\\ p=0.000\end{tabular}}           & \cellcolor[HTML]{FD6864}\textbf{\begin{tabular}[c]{@{}l@{}}-0.368\\ CI={[}-0.536,-0.172{]}\\ p=0.000\end{tabular}} & \begin{tabular}[c]{@{}l@{}}0.172\\ CI={[}-0.039,0.368{]}\\ p=0.110\end{tabular}                                    & \cellcolor[HTML]{FD6864}\textbf{\begin{tabular}[c]{@{}l@{}}-0.443\\ CI={[}-0.597,-0.258{]}\\ p=0.000\end{tabular}} \\ \hline
                             & Polarization                                                    & \begin{tabular}[c]{@{}l@{}}0.117\\ CI={[}-0.095,0.318{]}\\ p=0.279\end{tabular}                                              & \begin{tabular}[c]{@{}l@{}}0.102\\ CI={[}-0.110,0.305{]}\\ p=0.345\end{tabular}                                    & \begin{tabular}[c]{@{}l@{}}-0.132\\ CI={[}-0.332,0.080{]}\\ p=0.221\end{tabular}                                   & \begin{tabular}[c]{@{}l@{}}0.011\\ CI={[}-0.198,0.220{]}\\ p=0.916\end{tabular}                                    \\
                             & Volume                                                          & \cellcolor[HTML]{FD6864}\begin{tabular}[c]{@{}l@{}}-0.202\\ CI={[}-0.395,0.008{]}\\ p=0.059\end{tabular}                     & \cellcolor[HTML]{FD6864}\textit{\begin{tabular}[c]{@{}l@{}}-0.306\\ CI={[}-0.485,-0.104{]}\\ p=0.004\end{tabular}} & \cellcolor[HTML]{32CB00}\textbf{\begin{tabular}[c]{@{}l@{}}0.462\\ CI={[}0.280,0.612{]}\\ p=0.000\end{tabular}}    & \cellcolor[HTML]{FD6864}\begin{tabular}[c]{@{}l@{}}-0.267\\ CI={[}-0.452,-0.061{]}\\ p=0.012\end{tabular}          \\
                             & \% Volume                                                       & \cellcolor[HTML]{FD6864}\textbf{\begin{tabular}[c]{@{}l@{}}-0.688\\ CI={[}-0.784,-0.559{]}\\ p=0.000\end{tabular}}           & \cellcolor[HTML]{FD6864}\textbf{\begin{tabular}[c]{@{}l@{}}-0.689\\ CI={[}-0.785,-0.561{]}\\ p=0.000\end{tabular}} & \cellcolor[HTML]{32CB00}\textbf{\begin{tabular}[c]{@{}l@{}}0.615\\ CI={[}0.465,0.730{]}\\ p=0.000\end{tabular}}    & \cellcolor[HTML]{FD6864}\textbf{\begin{tabular}[c]{@{}l@{}}-0.804\\ CI={[}-0.867,-0.715{]}\\ p=0.000\end{tabular}} \\
\multirow{-4}{*}{Mask}       & \begin{tabular}[c]{@{}l@{}}Weighted\\ Polarization\end{tabular} & \cellcolor[HTML]{FD6864}\textbf{\begin{tabular}[c]{@{}l@{}}-0.687\\ CI={[}-0.783,-0.557{]}\\ p=0.000\end{tabular}}           & \cellcolor[HTML]{FD6864}\textbf{\begin{tabular}[c]{@{}l@{}}-0.688\\ CI={[}-0.784,-0.559{]}\\ p=0.000\end{tabular}} & \cellcolor[HTML]{32CB00}\textbf{\begin{tabular}[c]{@{}l@{}}0.610\\ CI={[}0.459,0.726{]}\\ p=0.000\end{tabular}}    & \cellcolor[HTML]{FD6864}\textbf{\begin{tabular}[c]{@{}l@{}}-0.804\\ CI={[}-0.867,-0.715{]}\\ p=0.000\end{tabular}} \\ \hline
                             & Polarization                                                    & \cellcolor[HTML]{32CB00}\begin{tabular}[c]{@{}l@{}}0.271\\ CI={[}0.066,0.455{]}\\ p=0.011\end{tabular}                       & \cellcolor[HTML]{32CB00}\textit{\begin{tabular}[c]{@{}l@{}}0.361\\ CI={[}0.164,0.530{]}\\ p=0.001\end{tabular}}    & \begin{tabular}[c]{@{}l@{}}-0.202\\ CI={[}-0.395,0.007{]}\\ p=0.059\end{tabular}                                   & \cellcolor[HTML]{32CB00}\begin{tabular}[c]{@{}l@{}}0.254\\ CI={[}0.047,0.440{]}\\ p=0.017\end{tabular}             \\
                             & Volume                                                          & \cellcolor[HTML]{32CB00}\textbf{\begin{tabular}[c]{@{}l@{}}0.718\\ CI={[}0.599,0.806{]}\\ p=0.000\end{tabular}}              & \cellcolor[HTML]{32CB00}\textbf{\begin{tabular}[c]{@{}l@{}}0.658\\ CI={[}0.520,0.762{]}\\ p=0.000\end{tabular}}    & \cellcolor[HTML]{FD6864}\textbf{\begin{tabular}[c]{@{}l@{}}-0.364\\ CI={[}-0.533,-0.168{]}\\ p=0.000\end{tabular}} & \cellcolor[HTML]{32CB00}\textbf{\begin{tabular}[c]{@{}l@{}}0.711\\ CI={[}0.590,0.801{]}\\ p=0.000\end{tabular}}    \\
                             & \% Volume                                                       & \cellcolor[HTML]{32CB00}\textbf{\begin{tabular}[c]{@{}l@{}}0.683\\ CI={[}0.553,0.781{]}\\ p=0.000\end{tabular}}              & \cellcolor[HTML]{32CB00}\textbf{\begin{tabular}[c]{@{}l@{}}0.672\\ CI={[}0.538,0.773{]}\\ p=0.000\end{tabular}}    & \cellcolor[HTML]{FD6864}\textbf{\begin{tabular}[c]{@{}l@{}}-0.532\\ CI={[}-0.667,-0.363{]}\\ p=0.000\end{tabular}} & \cellcolor[HTML]{32CB00}\textbf{\begin{tabular}[c]{@{}l@{}}0.781\\ CI={[}0.684,0.851{]}\\ p=0.000\end{tabular}}    \\
\multirow{-4}{*}{Vaccine}    & \begin{tabular}[c]{@{}l@{}}Weighted\\ Polarization\end{tabular} & \cellcolor[HTML]{32CB00}\textbf{\begin{tabular}[c]{@{}l@{}}0.687\\ CI={[}0.558,0.784{]}\\ p=0.000\end{tabular}}              & \cellcolor[HTML]{32CB00}\textbf{\begin{tabular}[c]{@{}l@{}}0.678\\ CI={[}0.546,0.777{]}\\ p=0.000\end{tabular}}    & \cellcolor[HTML]{FD6864}\textbf{\begin{tabular}[c]{@{}l@{}}-0.533\\ CI={[}-0.668,-0.364{]}\\ p=0.000\end{tabular}} & \cellcolor[HTML]{32CB00}\textbf{\begin{tabular}[c]{@{}l@{}}0.782\\ CI={[}0.684,0.852{]}\\ p=0.000\end{tabular}}    \\ \hline
                             & Sum                                                             & \begin{tabular}[c]{@{}l@{}}-0.121\\ CI={[}-0.322,0.091{]}\\ p=0.261\end{tabular}                                             & \begin{tabular}[c]{@{}l@{}}-0.044\\ CI={[}-0.251,0.167{]}\\ p=0.681\end{tabular}                                   & \begin{tabular}[c]{@{}l@{}}0.023\\ CI={[}-0.187,0.231{]}\\ p=0.831\end{tabular}                                    & \cellcolor[HTML]{FD6864}\textit{\begin{tabular}[c]{@{}l@{}}-0.315\\ CI={[}-0.492,-0.113{]}\\ p=0.003\end{tabular}} \\
\multirow{-2}{*}{Aggregated} & \begin{tabular}[c]{@{}l@{}}Weighted\\ Sum\end{tabular}          & \begin{tabular}[c]{@{}l@{}}-0.062\\ CI={[}-0.268,0.149{]}\\ p=0.566\end{tabular}                                             & \begin{tabular}[c]{@{}l@{}}0.027\\ CI={[}-0.183,0.236{]}\\ p=0.799\end{tabular}                                    & \begin{tabular}[c]{@{}l@{}}-0.032\\ CI={[}-0.240,0.178{]}\\ p=0.764\end{tabular}                                   & \begin{tabular}[c]{@{}l@{}}-0.256\\ CI={[}-0.442,-0.049{]}\\ p=0.016\end{tabular}                                 
\end{tabular}
\end{table}

\begin{table}[ht!]
\captionsetup{width=\linewidth}
\caption{Correlation Matrix between Topic Polarization and External Data in the United States. Bolded means $p < 0.001$. Italacized means $p < 0.01$. Underline means $p < 0.05$. Background color of green or red signifies the positive or negative correlation for significant p-values only.}
\label{table:us_corr_matrix}
\begin{tabular}{ll|llll}
                             &                                                                 & Cases                                                                                                              & Deaths                                                                                                             & Conspiracy (Volume)                                                                                             & Stringency Index                                                                                                   \\ \hline
                             & Polarization                                                    & \begin{tabular}[c]{@{}l@{}}-0.020\\ CI={[}-0.244,0.207{]}\\ p=0.867\end{tabular}                                   & \begin{tabular}[c]{@{}l@{}}-0.070\\ CI={[}-0.291,0.158{]}\\ p=0.548\end{tabular}                                   & \begin{tabular}[c]{@{}l@{}}-0.146\\ CI={[}-0.360,0.082{]}\\ p=0.207\end{tabular}                                & \begin{tabular}[c]{@{}l@{}}0.010\\ CI={[}-0.216,0.235{]}\\ p=0.931\end{tabular}                                    \\
                             & Volume                                                          & \begin{tabular}[c]{@{}l@{}}-0.196\\ CI={[}-0.403,0.031{]}\\ p=0.090\end{tabular}                                   & \cellcolor[HTML]{FD6864}\textit{\begin{tabular}[c]{@{}l@{}}-0.319\\ CI={[}-0.508,-0.100{]}\\ p=0.005\end{tabular}} & \cellcolor[HTML]{32CB00}\textbf{\begin{tabular}[c]{@{}l@{}}0.497\\ CI={[}0.306,0.650{]}\\ p=0.000\end{tabular}} & \cellcolor[HTML]{FD6864}\begin{tabular}[c]{@{}l@{}}-0.262\\ CI={[}-0.460,-0.038{]}\\ p=0.022\end{tabular}          \\
                             & \% Volume                                                       & \cellcolor[HTML]{FD6864}\textbf{\begin{tabular}[c]{@{}l@{}}-0.394\\ CI={[}-0.569,-0.185{]}\\ p=0.000\end{tabular}} & \cellcolor[HTML]{FD6864}\textbf{\begin{tabular}[c]{@{}l@{}}-0.436\\ CI={[}-0.602,-0.233{]}\\ p=0.000\end{tabular}} & \begin{tabular}[c]{@{}l@{}}0.120\\ CI={[}-0.109,0.336{]}\\ p=0.303\end{tabular}                                 & \cellcolor[HTML]{FD6864}\textit{\begin{tabular}[c]{@{}l@{}}-0.356\\ CI={[}-0.538,-0.142{]}\\ p=0.002\end{tabular}} \\
\multirow{-4}{*}{Lockdown}   & \begin{tabular}[c]{@{}l@{}}Weighted\\ Polarization\end{tabular} & \cellcolor[HTML]{FD6864}\textbf{\begin{tabular}[c]{@{}l@{}}-0.395\\ CI={[}-0.570,-0.186{]}\\ p=0.000\end{tabular}} & \cellcolor[HTML]{FD6864}\textbf{\begin{tabular}[c]{@{}l@{}}-0.442\\ CI={[}-0.607,-0.240{]}\\ p=0.000\end{tabular}} & \begin{tabular}[c]{@{}l@{}}0.101\\ CI={[}-0.128,0.319{]}\\ p=0.387\end{tabular}                                 & \cellcolor[HTML]{FD6864}\textit{\begin{tabular}[c]{@{}l@{}}-0.354\\ CI={[}-0.537,-0.140{]}\\ p=0.002\end{tabular}} \\ \hline
                             & Polarization                                                    & \begin{tabular}[c]{@{}l@{}}-0.103\\ CI={[}-0.321,0.125{]}\\ p=0.376\end{tabular}                                   & \begin{tabular}[c]{@{}l@{}}-0.108\\ CI={[}-0.325,0.121{]}\\ p=0.355\end{tabular}                                   & \begin{tabular}[c]{@{}l@{}}-0.078\\ CI={[}-0.298,0.150{]}\\ p=0.502\end{tabular}                                & \begin{tabular}[c]{@{}l@{}}-0.121\\ CI={[}-0.337,0.107{]}\\ p=0.298\end{tabular}                                   \\
                             & Volume                                                          & \cellcolor[HTML]{FD6864}\textit{\begin{tabular}[c]{@{}l@{}}-0.311\\ CI={[}-0.501,-0.092{]}\\ p=0.006\end{tabular}} & \cellcolor[HTML]{FD6864}\textit{\begin{tabular}[c]{@{}l@{}}-0.378\\ CI={[}-0.556,-0.167{]}\\ p=0.001\end{tabular}} & \cellcolor[HTML]{32CB00}\textbf{\begin{tabular}[c]{@{}l@{}}0.500\\ CI={[}0.310,0.652{]}\\ p=0.000\end{tabular}} & \cellcolor[HTML]{FD6864}\begin{tabular}[c]{@{}l@{}}-0.288\\ CI={[}-0.482,-0.067{]}\\ p=0.012\end{tabular}          \\
                             & \% Volume                                                       & \cellcolor[HTML]{FD6864}\textbf{\begin{tabular}[c]{@{}l@{}}-0.541\\ CI={[}-0.683,-0.360{]}\\ p=0.000\end{tabular}} & \cellcolor[HTML]{FD6864}\textbf{\begin{tabular}[c]{@{}l@{}}-0.526\\ CI={[}-0.672,-0.341{]}\\ p=0.000\end{tabular}} & \begin{tabular}[c]{@{}l@{}}-0.012\\ CI={[}-0.237,0.214{]}\\ p=0.915\end{tabular}                                & \cellcolor[HTML]{FD6864}\textbf{\begin{tabular}[c]{@{}l@{}}-0.452\\ CI={[}-0.615,-0.252{]}\\ p=0.000\end{tabular}} \\
\multirow{-4}{*}{Mask}       & \begin{tabular}[c]{@{}l@{}}Weighted\\ Polarization\end{tabular} & \cellcolor[HTML]{FD6864}\textbf{\begin{tabular}[c]{@{}l@{}}-0.565\\ CI={[}-0.701,-0.390{]}\\ p=0.000\end{tabular}} & \cellcolor[HTML]{FD6864}\textbf{\begin{tabular}[c]{@{}l@{}}-0.547\\ CI={[}-0.688,-0.367{]}\\ p=0.000\end{tabular}} & \begin{tabular}[c]{@{}l@{}}-0.020\\ CI={[}-0.245,0.206{]}\\ p=0.861\end{tabular}                                & \cellcolor[HTML]{FD6864}\textbf{\begin{tabular}[c]{@{}l@{}}-0.474\\ CI={[}-0.632,-0.279{]}\\ p=0.000\end{tabular}} \\ \hline
                             & Polarization                                                    & \cellcolor[HTML]{FD6864}\begin{tabular}[c]{@{}l@{}}-0.246\\ CI={[}-0.446,-0.021{]}\\ p=0.033\end{tabular}          & \begin{tabular}[c]{@{}l@{}}-0.216\\ CI={[}-0.421,0.010{]}\\ p=0.061\end{tabular}                                   & \cellcolor[HTML]{FD6864}\begin{tabular}[c]{@{}l@{}}-0.268\\ CI={[}-0.466,-0.046{]}\\ p=0.019\end{tabular}       & \begin{tabular}[c]{@{}l@{}}-0.133\\ CI={[}-0.348,0.096{]}\\ p=0.253\end{tabular}                                   \\
                             & Volume                                                          & \cellcolor[HTML]{32CB00}\textit{\begin{tabular}[c]{@{}l@{}}0.360\\ CI={[}0.147,0.542{]}\\ p=0.001\end{tabular}}    & \cellcolor[HTML]{32CB00}\textit{\begin{tabular}[c]{@{}l@{}}0.305\\ CI={[}0.085,0.496{]}\\ p=0.007\end{tabular}}    & \cellcolor[HTML]{32CB00}\textit{\begin{tabular}[c]{@{}l@{}}0.303\\ CI={[}0.084,0.495{]}\\ p=0.008\end{tabular}} & \begin{tabular}[c]{@{}l@{}}0.196\\ CI={[}-0.031,0.404{]}\\ p=0.089\end{tabular}                                    \\
                             & \% Volume                                                       & \cellcolor[HTML]{32CB00}\textbf{\begin{tabular}[c]{@{}l@{}}0.666\\ CI={[}0.518,0.775{]}\\ p=0.000\end{tabular}}    & \cellcolor[HTML]{32CB00}\textbf{\begin{tabular}[c]{@{}l@{}}0.682\\ CI={[}0.539,0.786{]}\\ p=0.000\end{tabular}}    & \begin{tabular}[c]{@{}l@{}}-0.068\\ CI={[}-0.289,0.160{]}\\ p=0.561\end{tabular}                                & \cellcolor[HTML]{32CB00}\textbf{\begin{tabular}[c]{@{}l@{}}0.573\\ CI={[}0.400,0.707{]}\\ p=0.000\end{tabular}}    \\
\multirow{-4}{*}{Vaccine}    & \begin{tabular}[c]{@{}l@{}}Weighted\\ Polarization\end{tabular} & \cellcolor[HTML]{32CB00}\textbf{\begin{tabular}[c]{@{}l@{}}0.633\\ CI={[}0.475,0.751{]}\\ p=0.000\end{tabular}}    & \cellcolor[HTML]{32CB00}\textbf{\begin{tabular}[c]{@{}l@{}}0.655\\ CI={[}0.505,0.767{]}\\ p=0.000\end{tabular}}    & \begin{tabular}[c]{@{}l@{}}-0.115\\ CI={[}-0.332,0.113{]}\\ p=0.322\end{tabular}                                & \cellcolor[HTML]{32CB00}\textbf{\begin{tabular}[c]{@{}l@{}}0.576\\ CI={[}0.403,0.710{]}\\ p=0.000\end{tabular}}    \\ \hline
                             & Sum                                                             & \begin{tabular}[c]{@{}l@{}}-0.184\\ CI={[}-0.393,0.044{]}\\ p=0.112\end{tabular}                                   & \begin{tabular}[c]{@{}l@{}}-0.196\\ CI={[}-0.403,0.031{]}\\ p=0.090\end{tabular}                                   & \cellcolor[HTML]{FD6864}\begin{tabular}[c]{@{}l@{}}-0.247\\ CI={[}-0.448,-0.023{]}\\ p=0.031\end{tabular}       & \begin{tabular}[c]{@{}l@{}}-0.119\\ CI={[}-0.335,0.110{]}\\ p=0.307\end{tabular}                                   \\
\multirow{-2}{*}{Aggregated} & \begin{tabular}[c]{@{}l@{}}Weighted\\ Sum\end{tabular}          & \begin{tabular}[c]{@{}l@{}}-0.217\\ CI={[}-0.422,0.009{]}\\ p=0.060\end{tabular}                                   & \begin{tabular}[c]{@{}l@{}}-0.198\\ CI={[}-0.405,0.029{]}\\ p=0.086\end{tabular}                                   & \begin{tabular}[c]{@{}l@{}}-0.193\\ CI={[}-0.401,0.034{]}\\ p=0.095\end{tabular}                                & \begin{tabular}[c]{@{}l@{}}-0.091\\ CI={[}-0.310,0.137{]}\\ p=0.434\end{tabular}                                  
\end{tabular}
\end{table}

\subsection{The Relationship between National Partisan Polarization and Epidemiological Data}
\begin{figure}[ht!]
    \begin{subfigure}{0.48\linewidth}
        \centering
        \includegraphics[scale=0.45,trim={0 0 0 .75cm},clip]{Daily_Polarization/CAD_political/unweighted_aggregated_poli.png}
        \caption{Canada: Daily Polarization Trends}
        \label{fig:cad_political_aggregated_plot}
    \end{subfigure}
    \hspace{0.01\linewidth}
    \begin{subfigure}{0.48\linewidth}
        \centering
        \includegraphics[scale=0.43,trim={0 0 0 .75cm},clip]{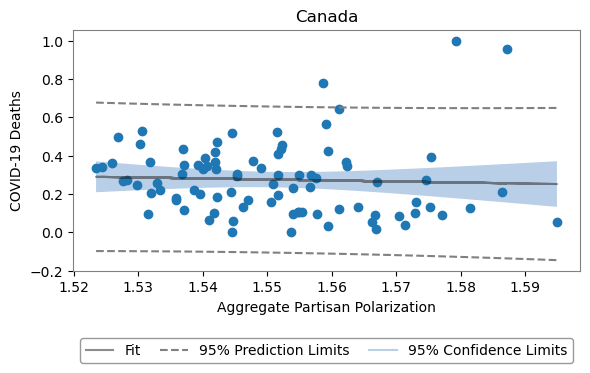}
        \caption{Canada: Polarization v.s. Deaths}
        \label{fig:cad_political_aggregated_corr_deaths}
    \end{subfigure}
    \\
    \begin{subfigure}{0.48\linewidth}
        \centering
        \includegraphics[scale=0.45,trim={0 0 0 .75cm},clip]{Daily_Polarization/US/unweighted_aggregated_poli.png}
        \caption{United States: Daily Polarization Trends}
        \label{fig:us_aggregated_plot}
    \end{subfigure}
    \hspace{0.01\linewidth}
    \begin{subfigure}{0.48\linewidth}
        \centering
        \includegraphics[scale=0.43,trim={0 0 0 .75cm},clip]{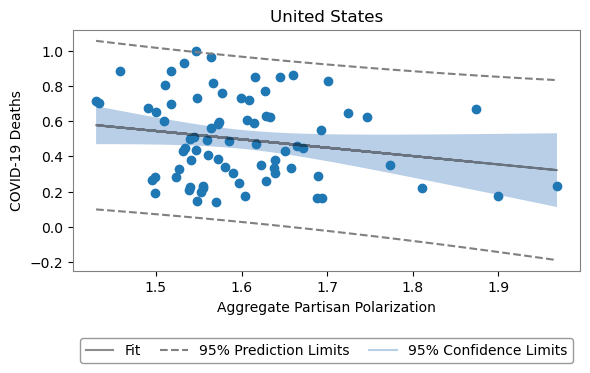}
        \caption{United States:  Polarization v.s. Deaths}
        \label{fig:us_aggregated_corr_deaths}
    \end{subfigure}
    \caption{Daily aggregate polarization v.s. \COVID new cases and deaths for (a) Canada and (c) the United States from October \nth{9}, 2020 to January \nth{3}, 2021. Correlation of polarization and \COVID deaths for (b) Canada and (d) the United States. The correlation coefficient are \textbf{-0.044} for Canada with CI = [0.251,0.167] (n = 88, p = 0.681) and \textbf{-0.196} for the United States with  CI = [-0.403,0.031] (n = 88, p = 0.090).}
    \label{fig:Aggregated_corr}
\end{figure}

Here, we investigate the aggregated polarization over time for each country and how it relates to the reported number of New Cases and Deaths for \COVID.
In Figure \ref{fig:Aggregated_corr}, we observe\textit{ that polarization is not correlated with the severity of the pandemic, in both  Canada and the United States.}  
To compute the daily aggregate polarization measure, we employ a weighted sum of each topic's polarization, considering the percentage of each topic's tweets within that day's volume of \COVID-related tweets.

\section{Methodology Details \label{sec:methodology_cont}}

In the following section, we report the classification metrics of each module in the pipeline for measuring polarization.

\subsection{Classifying Tweets by Topics}
\begin{table}[th!]
\captionsetup{width=\linewidth}
\caption{Tweet Topic Classification Metrics. 200 relevant and 200 irrelevant tweets were sampled for manual annotation. F1-score is calculated on the true labels where both annotators agreed upon over 5 runs with a different random seed.}
\label{table:Tweet_Classification_Statistics}
\centering
\begin{tabular}{l|r r r r r}
\multicolumn{6}{c}{Canada} \\
Topic & Relevant & Irrelevant & Cohen & F1-Score & \# of Tweets \\
\hline
 Lockdown   & 170 & 356 & 0.73 & 97.13 ± 1.57 & 1,553,984  \\
 Mask       & 292 & 282 & 0.91 & 98.48 ± 1.24 & 1,994,293  \\
 Vaccine    & 326 & 248 & 0.91 & 99.84 ± 0.31 & 2,145,549  \\
 Conspiracy & 338 & 171 & 0.67 & 97.20 ± 0.71 & 16,575,934 \\
 \\
\multicolumn{6}{c}{United States} \\
Topic & Relevant & Irrelevant & Cohen & F1-Score & \# of Tweets \\
\hline
 Lockdown   & 126 & 199 & 0.63 & 100.00 ± 0.00 & 897,565 \\
 Mask       & 197 & 200 & 0.98 & 99.49 ± 0.62  & 1,562,706 \\
 Vaccine    & 192 & 201 & 0.96 & 100.00 ± 0.00 & 1,541,360 \\
 Conspiracy & 195 & 155 & 0.75 & 95.55 ± 2.39  & 926,389 \\
\end{tabular}
\end{table}

\subsection{Classifying Users by Geolocation}

\begin{table}[th!]
\captionsetup{width=\linewidth}
\caption{Geolocated Users Number and Correlation. Correlation is done with the official 2021 population census for each country.}
\label{table:Geolocation_correlation}
\centering
\begin{tabular}{l|ll}

\multicolumn{3}{c}{Canada} \\
Users & Total & Correlation \\
\hline
Geolocated & 282,454 & \textbf{0.92} (n=13, p=6.20e-06, CI=[0.76, 0.98])  \\ 
w/ Party Affiliation & 195,456 & \textbf{0.92} (n=13, p=6.83e-06, CI=[0.76, 0.98]) \\

\\

\multicolumn{3}{c}{United States} \\
Users & Total & Correlation \\
\hline
Geolocated & 757,601 & \textbf{0.98} (n=52, p=9.27e-35, CI=[0.96, 0.99]) \\
w/ Party Affiliation & 242,056 & \textbf{0.97} (n=52, p=2.51e-34, CI=[0.96, 0.99]) \\

\end{tabular}
\end{table}

\subsection{Classifying User Party Affiliation}

We report the classification metrics for Canada in Table \ref{table:CAD_UPA_Classification_Accuracy} and for the United States in Table \ref{table:US_UPA_Classification_Accuracy}. For Canada, our model classified users for each parties for the activity as well, but the F1-score was not satisfactory as parties within the liberal (left) party family and conservative (right) party family was easily confused as shown in the confusion matrix in Table \ref{table:CAD_UPA_CNFSN_MTRX}. Thus, for our analysis, we merged the parties in Canada, and show the confusion matrix for after the merge in Table \ref{table:CAD_UPA_CNFSN_MTRX_SMPL}.

\begin{table}[ht!]
\captionsetup{width=\linewidth}
\caption{Canadian User Party Affiliation Classification. Cohen Kappa score of 0.74}
\label{table:CAD_UPA_Classification_Accuracy}
\centering
\begin{tabular}{ll|rrr}
 & Party & Support & F1-Score & \# of Users \\
 \hline
\multirow{6}{*}{Profile} & CPC &  98 & 92.93 ± 1.12 &  1,769 \\
                         & GPC &  60 & 88.50 ± 1.54 &  97 \\
                         & LPC &  100 & 90.89 ± 1.34 & 783 \\
                         & NDP &  124 & 93.44 ± 0.53 & 370 \\
                         & NO\_PARTY & 105 & 86.16 ± 1.97 & 667 \\
                         & PPC & 95 & 94.09 ± 1.34 & 402 \\
  \hline
\multirow{2}{*}{Activity} & RPF & 628 & 93.85 ± 0.47 & 193,225 \\
                          & LPF & 357 & 89.10 ± 0.73 & 299,836 \\
  \hline
\multirow{2}{*}{Combined} & RPF & -- & -- & 196,338 \\
                          & LPF & -- & -- & 302,023
\end{tabular}
\end{table}

\begin{table}[ht!]
\captionsetup{width=\linewidth}
\caption{Canadian User Party Affiliation Activity Classifier Confusion Matrix}
\label{table:CAD_UPA_CNFSN_MTRX}
\centering
\begin{tabular}{l|lllll}
 & CPC & PPC & GPC & LPC & NDP \\
 \hline
CPC & 398 & 24 & 1 & 29 & 6 \\
PPC & 59 & 49 & 0 & 1 & 1 \\
GPC & 3 & 0 & 6 & 10 & 6 \\
LPC & 18 & 3 & 2 & 161 & 26 \\
NDP & 5 & 1 & 1 & 19 & 58
\end{tabular}
\end{table}

\begin{table}[ht!]
\captionsetup{width=\linewidth}
\caption{Canadian User Party Affiliation Activity Classifier Confusion Matrix - Liberal (left) and Conservative (right) Party Family}
\label{table:CAD_UPA_CNFSN_MTRX_SMPL}
\centering
\begin{tabular}{l|cc}
 & Right Party Family & Left Party Family \\
 \hline
Right Party Family & 530 & 38 \\
Left Party Family & 27 & 289
\end{tabular}
\end{table}

\begin{table}[ht!]
\captionsetup{width=\linewidth}
\caption{American User Party Affiliation Classification. Cohen Kappa score of 0.76. Total 763,164 users.}
\label{table:US_UPA_Classification_Accuracy}
\centering
\begin{tabular}{ll|rrr}
 & Party & Support & F1-Score & \# of Users \\
 \hline
\multirow{2}{*}{Profile} & Republican & 854 & 97.21 ± 0.66 & 86,989 \\
                         & Democrat   & 928 & 97.40 ± 0.63 & 82,923 \\
  \hline
\multirow{2}{*}{Activity} & Republican & 10,583 & 92.98 ± 0.18 & 239,449 \\
                          & Democrat   & 10,226 & 92.99 ± 0.16 & 145,733 \\
  \hline
\multirow{2}{*}{Combined} & Republican & -- & -- & 336,231 \\
                          & Democrat   & -- & -- & 426,933
\end{tabular}
\end{table}

\subsection{Distribution Matching with Election Results}

\begin{figure}[ht!]\centering
    \begin{subfigure}{0.44\linewidth}
        \centering
        \includegraphics[scale=0.45]{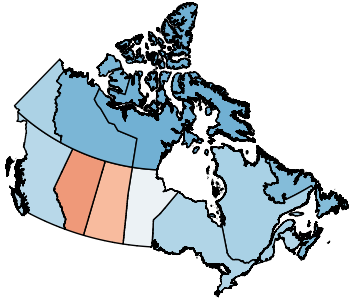}
        \caption{2019 Election Results}
        \label{fig:cad_pol_election_results}
    \end{subfigure}\hspace{0.05\linewidth}
    \begin{subfigure}{0.44\linewidth}
        \centering
        \includegraphics[scale=0.45]{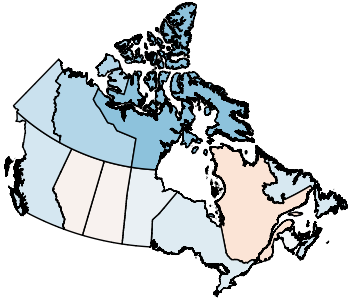}
        \caption{Geolocated Users}
        \label{fig:cad_pol_aggregate_geolocated}
    \end{subfigure}\\

   \begin{subfigure}{0.44\linewidth}
        \centering
        \includegraphics[scale=0.45]{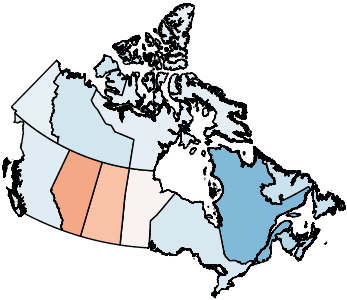}
        \caption{Lockdown}
        \label{fig:cad_pol_aggregate_lockdown}
    \end{subfigure}\hspace{0.05\linewidth}
    \begin{subfigure}{0.44\linewidth}
        \centering
        \includegraphics[scale=0.45]{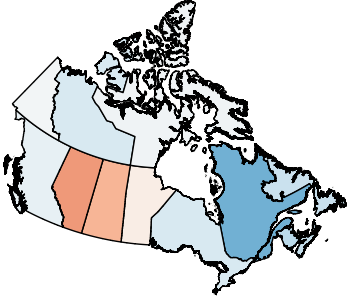}
        \caption{Mask}
        \label{fig:cad_pol_aggregate_mask}
    \end{subfigure}\\

    \begin{subfigure}{0.44\linewidth}
        \centering
        \includegraphics[scale=0.45]{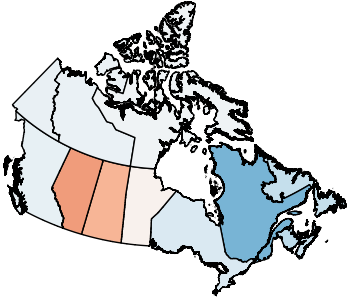}
        \caption{Vaccine}
        \label{fig:cad_pol_aggregate_vaccine}
    \end{subfigure}\hspace{0.05\linewidth}
    \begin{subfigure}{0.44\linewidth}
        \centering
        \includegraphics[scale=0.45]{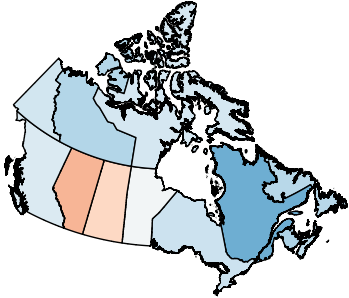}
        \caption{Conspiracy}
        \label{fig:cad_pol_aggregate_conspiracy}
    \end{subfigure}\\
   
    \caption{Normalized distribution of the inferred user party affiliations compared to the CAD 2021 election results.}
    \label{fig:cad_pol_aggregate_party_affliation}
\end{figure}

\begin{figure}[ht!]\centering
    \begin{subfigure}{0.44\linewidth}
        \centering
        \includegraphics[scale=0.45]{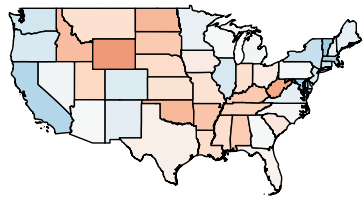}
        \caption{2020 Election Results}
        \label{fig:us_election_results}
    \end{subfigure}\hspace{0.05\linewidth}
     \begin{subfigure}{0.44\linewidth}
        \centering
        \includegraphics[scale=0.45]{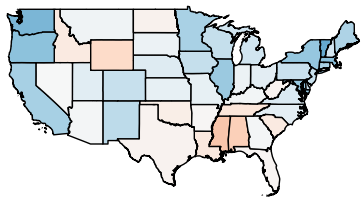}
        \caption{Geolocated Users}
        \label{fig:us_aggregate_geolocated}
    \end{subfigure}\\

   \begin{subfigure}{0.44\linewidth}
        \centering
        \includegraphics[scale=0.45]{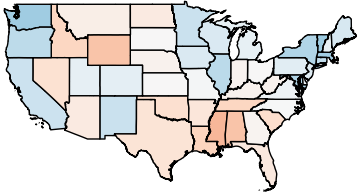}
        \caption{Lockdown}
        \label{fig:us_aggregate_lockdown}
    \end{subfigure}\hspace{0.05\linewidth}
    \begin{subfigure}{0.44\linewidth}
        \centering
        \includegraphics[scale=0.45]{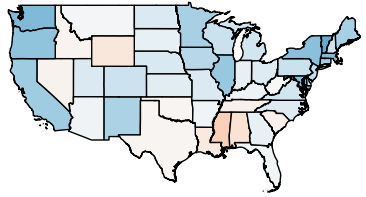}
        \caption{Mask}
        \label{fig:us_aggregate_mask}
    \end{subfigure}\\

    \begin{subfigure}{0.44\linewidth}
        \centering
        \includegraphics[scale=0.45]{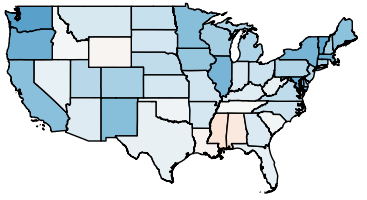}
        \caption{Vaccine}
        \label{fig:us_aggregate_vaccine}
    \end{subfigure}\hspace{0.05\linewidth}
    \begin{subfigure}{0.44\linewidth}
        \centering
        \includegraphics[scale=0.45]{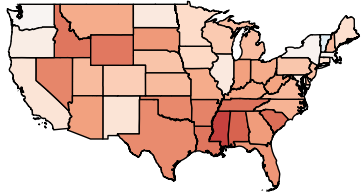}
        \caption{Conspiracy}
        \label{fig:us_aggregate_conspiracy}
    \end{subfigure}\\

    \caption{Empirical distribution of the inferred user party affiliations compared to the US 2020 election results. Interestingly, lockdown matches closer to the election results, mask and vaccine has a higher liberal ratio and conspiracy has a higher conservative ratio.}
    \label{fig:us_aggregate_party_affliation}
\end{figure}

We further validate our party affiliation distribution using the 2019 Canada Federal Election and the US 2020 Election results for the respective country. We calculate the correlation between the ratio of numbers of liberal and conservative families-labelled users per region in our data compared to the election results and obtain strong correlations of \textbf{0.815} for Canada visualized in Figure \ref{fig:cad_pol_aggregate_party_affliation} and \textbf{0.802} for the United States visualized in Figure \ref{fig:us_aggregate_party_affliation}. We also visualize the ratio for all geolocated users for the respective topics.

\subsection{Matching Users to US Voter Registration}
Following the best practice for evaluating party affiliation predictions \cite{barbera2015birds}, we matched Twitter (X) users from our dataset with the primary voter registration records available for five states: Ohio, New York, Florida, Arkansas, North Carolina, as well as Washington DC. From these records, we obtain the party affiliation of unique users in each state by md5-hashing their names and county to construct a key identifier. From our set of geolocated Twitter (X) users, we kept everyone that belonged to one of the five states or DC, and we removed those whose location could not be retrieved. Finally, we matched the most recent voter party affiliation records from the registration data to the unique Twitter (X) users that matched both the county and either the first name and last name or the first, middle and last name. We pre-processed the user's name on Twitter (X) to remove emojis. After matching, we removed users not affiliated with either of the two major parties and users whose name matched with more than one record per county (indicating a non-unique match). 

We then compare users' party from voter registration with their predicted party, first using their profile description and second their \COVID-related tweets.
Using our geolocation and voter record matching, Table \ref{table:voter_registration_stats} shows we are able match  a significant number of  users, more than 30k, across the 5 states and DC with their voter records.

\begin{table}[ht!]
    \caption{Users matched to their voter registration. Voters* and Users* respectively correspond to the number of unique voters in the records and unique Twitter (X) users in our data.}
    \centering
    \begin{tabular}{l|llllll}
        State                & Voters* & Users* & Matched & Democrat & Republican & Other \\ \hline
        Ohio                 & 7,771,590     & 4,913          &  1,431    & 320 & 193 & 917    \\
        New York             & 17,718,437    & 30,927             & 8,255  & 4,843 & 1,631 & 1,781       \\
        Florida              & 14,477,882    & 50,541             & 12,905  & 5,585 & 4,508 & 2,810      \\
        Arkansas             & 1,722,465     & 4,311              & 1,280   & 145 & 140 & 995      \\
        District of Columbia & 510,026       & 17,661             & 2,538   & 1,929 & 153 & 456      \\
        North Carolina       & 8,004,814     & 20,761             & 6,050    & 2,450 & 1,655 & 1,945  \\
        \hline
        Total & & & 32,456 & 15,272 & 8,280 & 8,904
    \label{table:voter_registration_stats}
    \end{tabular}

\end{table}

 We get an accuracy of 74.35\% for the first method and 73.35\% for the second one. We note that both methods are binary classifiers while users can also be independent or support a third party, despite an ideology (and behavior/voting) that aligns strongly with one of the two main parties. They can also have an outdated registration that no longer reflects the beliefs they currently hold and express on Twitter (X). Therefore, although accuracy here is lower, it still indicates our classification is acceptable and on par with the standard for this type of evaluation \citep{barbera2015birds}.

\subsection{Approximation of Polarization}

We test accuracy of approximation in Algorithm 1 over the daily Lockdown and Vaccine tweets. 

\begin{figure}[ht!]
    \begin{subfigure}{0.47\linewidth}
        \centering
        \includegraphics[height=0.75\columnwidth,width=0.99\columnwidth]{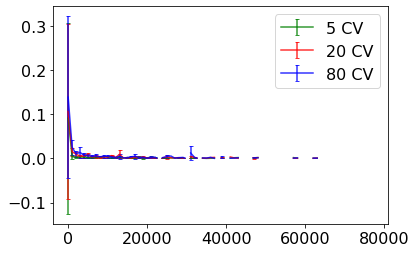}
        \caption{Approximation Quality}
        \label{fig:C_Index_Quality}
    \end{subfigure}
    \hspace{0.01\linewidth}
    \begin{subfigure}{0.47\linewidth}
        \centering
        \includegraphics[height=0.75\columnwidth,width=0.99\columnwidth]{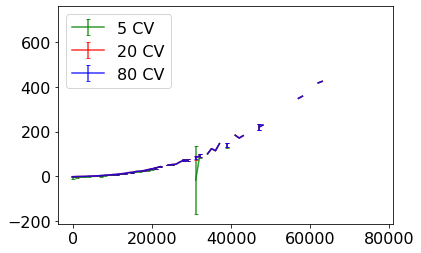}
        \caption{Time Saved in Seconds}
    \label{fig:C_Index_Time_Saved}
    \end{subfigure}
    \caption{
    For large enough data, the approximation error is close to 0.001. Our approximation of polarization also exponentially saves time as the number of instances grows.
    }
\end{figure}

In Figure \ref{fig:C_Index_Quality}, we plot the total number of users against the absolute error (and its standard deviation) of the approximated $poli$ compared to the exact value, binned for every 1,000 users. We observe a dramatic drop of the absolute error term at around 3,000 users. When we reach the 10,000 users value, the absolute error is usually below 0.001.

In Figure \ref{fig:C_Index_Time_Saved}, we plot the total number of users against the time saved in running the approximation algorithm compared to  running the exact $poli$, binned also for every 1,000 users. We observe that the time saved is exponential to the number of users. We note that at around 50,000 users, the approximation rarely needs to increase the fraction of users sampled.  

These findings confirm that we can accurately approximate $poli$ for large-scale data that is impossible to measure exactly because of memory constraints. As $poli$ relies heavily on finding pairwise distances (time and memory intensive), we see from our analysis how sampling can save both time and memory exponentially. 

\begin{figure}[ht!]\centering
    \begin{subfigure}{\linewidth}
        \centering
        \includegraphics[scale=0.6]{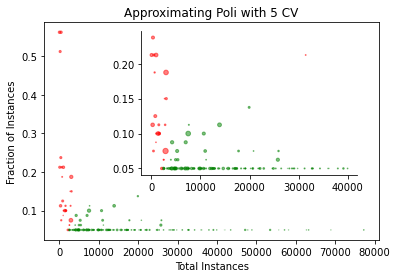}
        \label{fig:approximation_of_c_index_5}
    \end{subfigure}
    \\
    \begin{subfigure}{\linewidth}
        \centering
        \includegraphics[scale=0.6]{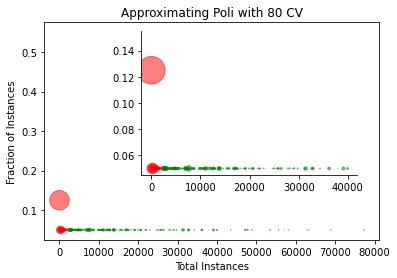}
        \label{fig:approximation_of_c_index_80}
    \end{subfigure}\\
    \caption{Approximation of $poli$. Green dots mean time was saved while red means time lost. The circle size is the absolute error times 1000. The inner plot zooms in and doubles the circle sizes for clarity. We observe that the fraction of instances needed increase as we decrease the needed minimum coefficient of variation. This follows our intuition as a larger sample of data is much easier to approximate the full data. We also that at lower minimum coefficient of variation, the less error we have (circles are much smaller). On all plots, we also see that we always save time (green circles) if there are more than 10,000 instances. 
    }
    \label{fig:approximation_of_c_index}
\end{figure}

\begin{figure}[ht!]\centering
    \begin{subfigure}{\linewidth}
        \centering
        \includegraphics[scale=0.6]{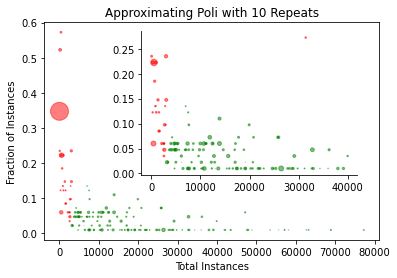}
        \label{fig:approximation_of_c_index_10_RC}
    \end{subfigure}
    \\
    \begin{subfigure}{\linewidth}
        \centering
        \includegraphics[scale=0.6]{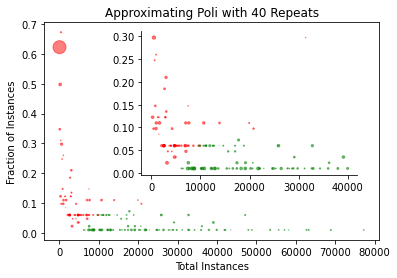}
        \label{fig:approximation_of_c_index_40_RC}
    \end{subfigure}\\
    \caption{Approximation of $poli$. 
    As in the preceding figure, we observe that as we increase the number of repeats, the less error we have in our approximation (smaller circles). Likewise the amount of time saved decreases as we increase the repeat count (amount of red circles increase as repeats increase). We also observe a general slight increase in the fraction of instances needed as we increase the repeats.}
    \label{fig:approximation_of_c_index_2}
\end{figure}

We also explore the impact of changing the minimum coefficient of variation. We start with a minimum sample size of 1\% or $fraction = 0.01$. We keep the step\_size constant at 0.01. For the first experiment, $repeats$ is set to $10$. For the second experiment, shown here in figures \ref{fig:approximation_of_c_index} and \ref{fig:approximation_of_c_index_2}, $epsilon$ is set to $0.05$.

\end{document}